\documentclass[12pt]{article}
\usepackage{amsmath,amssymb}
\usepackage[english]{babel}
\title{Vacuum engineering at a photon collider?}
\author{ E.~A.~Kuraev$^1$ and Z.~K.~Silagadze$^2$
\vspace*{3mm} \\
$^1$ Joint Institute for Nuclear Research, 141 980 Dubna, Russia \\
$^2$ Budker Institute of Nuclear Physics,  630 090 Novosibirsk, Russia }
\date{}

\begin{document}
\large
\maketitle

\begin{abstract}
The aim of this paper is twofold: to provide a rather detailed and 
self-contained introduction into the physics of the Disoriented Chiral 
Condensate (DCC) for the photon (and linear) collider community, and to 
indicate that such physics can be searched and studied at photon colliders.
Some side tracks are also occasionally followed during the exposition, if 
they lead to interesting vistas. For gourmets, the Baked Alaska recipe is 
given in the appendix.
\end{abstract}

\section{Introduction}
The twentieth century witnessed tremendous progress in our understanding 
of the fundamental building blocks of matter and their interactions. Not the 
least role in this success was played by continuous advance in accelerator
technologies. At the beginning of the new century, accelerator-based 
experiments are expected to preserve their leading role in the field of 
high-energy physics \cite{1:1}. 

Over the seven decades since Lowrence's first
cyclotron one has observed a nearly exponential growth in effective energies
of the accelerators by the increment factor of about 25 per decade 
(the Levingston law \cite{1:2}. By the effective energy one means the 
laboratory energy of particles colliding with a proton at rest to reach 
the same center of mass energy). At that the cost per unit effective energy
has decreased  by about four orders of magnitudes. This is indeed a remarkable
trend and it was fed by a succession of new ideas and technologies 
\cite{1:2}: the principle of phase stability, strong focusing, high impedance
microwave devices, superconducting technologies, storage rings and beam
cooling.

However the accelerators were becoming ever bigger and more expensive on 
the whole.
We have already entered ``the dinosaur era'' with monstrous machines and the
Levingston tendency is slowing its pace. The problem with the circular
$e^+e^-$ colliders is that the synchrotron radiation severely limits maximal
attainable energy. It is believed that this technology has reached its limits
at LEP and no other bigger project of this type will be ever realized. Instead
the linear $e^+e^-$ colliders are considered as a viable 
alternative. Extension of the existing linear accelerator technology towards 
higher accelerating gradients and smaller emittance beams is expected to make
real a design of the $\mathrm{TeV}$ scale linear colliders. Further progress
with the conventional techniques is problematic unless some radically new idea
appears. In fact the high gradient efficient acceleration is a tough thing.
In a free electromagnetic wave the $E$ field is at right angle to the particle
momentum and no efficient acceleration can be achieved. For efficient 
acceleration one has to have matter very near or within the beams. Then energy
considerations combined with the survivability of the accelerating structure 
limits the attainable acceleration gradient \cite{1:1,1:2}. 

The proton circular colliders still have some reserve left because, owing 
to the heaviness of the proton, the synchrotron radiation constraint is 
expected only at very high energies. The Large Hadron Collider (LHC) with 
$7~\mathrm{TeV}$ proton beams is under construction now. LHC is a very 
important
high-energy physics project and we believe that its results will determine the 
future shape of the field. An analogous collider with the center of mass energy
about $100~\mathrm{TeV}$ seems also feasible and maybe the Very Large Hadron 
Collider (VLHC) will be the last monstrous dinosaur of this type.

Other possibilities include the muon colliders first suggested by G.~I.~Budker
many years ago \cite{1:3}. Muons, being about 207 times heavier than 
electrons, experience much less radiative energy losses, which are inversely
proportional to the forth power of the particle mass. It seems that the 
efficient multi-$\mathrm{TeV}$ muon colliders can be constructed despite the
fact that the muon is an unstable particle \cite{1:3}.    

But why all the fuss? Are these future very complex and costly accelerators
really necessary? The past research led to the triumph of the Standard Model.
At that the revolutionary 70's were followed by decades of the more or less 
routine verification of the Standard Model wisdom -- the situation eloquently
expressed by Bjorken some time ago \cite{1:1}: ``a theorist working within the
Standard Model feels like an engineer, and one working beyond it feels like 
a crackpot''. Since then ``crackpots'' have developed a string theory as the
main
challenge to the standard paradigm \cite{1:4}. This ``Theory of Everything''
is full of deep and beautiful mathematical constructs and is generally
considered as the most promising road towards understanding fundamental
physics. The only trouble with it is that it will be extremely difficult to
check experimentally the predictions of this theory, because the most direct
predictions refer to the nature of space-time at the Plank scale, 
$\sim 10^{19}~\mathrm{GeV}$, and no experimental method seems to be ever 
able to access such energies in a foreseeable future. So the string theorists
are doomed to face the fatal question ``can there be physics without 
experiments?'' \cite{1:5} for a long time. Therefore, on the one hand, we have
a clear experimental and theoretical success up to the electroweak scale, 
$\sim 100~\mathrm{GeV}$, where the Standard Model reigns, and, on the other
hand one has a very ambitious theory without any clues how to check it
experimentally. But what lies in between, worth of billions of dollars 
to spend in future accelerators and detectors, to investigate?

Despite its splendid success, nobody doubts that the Standard Model will break
down sooner or later. There are several reasons why the Standard Model cannot
be the final theory and why some new physics beyond the Standard Model is 
expected \cite{1:6}:
\begin{itemize}
\item $SU(3)\times SU(2)\times U(1)$ symmetry group defines separate gauge 
theories with three different coupling constants. The conceptual similarity
of these theories is begging for unification.
\item The family problem -- why are there three quark-lepton families?
\item The origin of the quark and lepton masses and mixing angles, as well as
of the $CP$ violation.
\item Solid experimental evidence of the neutrino oscillations require 
non-vanishing neutrino masses and therefore some extension of the Standard 
Model. However, very minimal extension might be sufficient to accommodate
neutrino masses.
\item The strong $CP$ problem -- why is the allowed $CP$ violating 
$\theta$-term in the QCD Lagrangian very small or absent? 
\item The hierarchy problem -- why is the electroweak scale so different from
the Plank scale?
\item The cosmological constant problem -- why gravity almost does not feel
the presence of various symmetry-breaking condensates?
\end{itemize}
But how far is this expected new physics? The logical structure of the 
Standard Model itself hints that quite interesting and crucial things can 
happen in the realm of the next generation of the future colliders. One of the
main guiding principles of the Standard Model, which plays a key role in the 
theory, is gauge symmetry. The historical roots of the gauge invariance are
reviewed by Jackson and Okun \cite{1:7} and the review embraces about two
centuries. In fact one can go even further through history, another twenty
centuries or so up to the times of the ancient Greece, and find the roots in
the most widely known theorem from Euclid's ``Elements of Geometry'': The sum
of the interior angles of a triangle equals 180 degree\!. Euclid deduces this
theorem from the so-called parallel axiom. All efforts to avoid this 
sophisticated axiom failed and finally led to the discovery of non-Euclidean
geometry. But we do not follow this track. Instead, we start to generalize 
Euclid's 180 degree theorem step by step \cite{1:8}. The first step involves
the concept of exterior angle: the interior angle $\alpha$ and the 
corresponding exterior angle $\beta$ are related by $\alpha+\beta=\pi$. Then
the theorem immediately generalizes from triangles to arbitrary polygons: The
sum of the exterior angles of a polygon equals $2\pi$. 

Let us now consider a triangle whose edges are not straight lines but
some plane smooth curves. When the unit tangent vector is transported by a
length $\Delta l$ along the smooth curve, it turns through an angle $\Delta
\phi$. The limit of the ratio $\Delta \phi / \Delta l$, when $\Delta l\to
0$, defines the geodesic curvature of the curve. Therefore, for such a curved
triangle the 180 degree theorem takes the form
$$\sum \mathrm{ext. \;angles} +\int \mathrm{geod.\; curv.} =2\pi,$$
where the integral is along the triangle edges. This follows from the
fact that any curved triangle can be approximated by a polygon and then
the total turning of the tangent along the edges (the integral geodesic
curvature of the edges) is given by the sum of the corresponding exterior
angles.

One can define the geodesic curvature by using normals instead of tangents,
because the normal rotates exactly as the tangent does when a point moves
along the curve. The advantage of using normals is that one can generalize
the concept of curvature to surfaces which have no unique tangent direction
but the direction of the normal is still well defined. The corresponding
generalization is called the Gaussian curvature \cite{1:8} and the 180
degree theorem for a general triangle on a curved surface looks like
\begin{equation}
\sum \mathrm{ext.\; angles} +\int \mathrm{geod.\; curv.} 
+\iint \mathrm{Gaussian \; curv.} =2\pi.
\label{GBtriangle} \end{equation}

Finally, let $D$ be a domain on the surface whose boundary $\partial D$ is
formed by one or more sectionally-smooth curves. We can triangulate $D$ with
triangles which have geodesic inside (not belonging to $\partial D$) 
edges. For each triangle we will have (\ref{GBtriangle}). If we add these
equations up and rearrange the angles cleverly we get the Gauss-Bonnet formula
\cite{1:8}
\begin{equation}
\sum \mathrm{ext.\; angles} +\int\limits_{\partial D} \mathrm{geod.\; curv.} 
+\iint\limits_D \mathrm{Gaussian \; curv.} =2\pi\,\chi(D),
\label{GBtheorem} \end{equation}
where $\chi(D)=v-e+f$, $v$ being the number of vertices, $e$ -- the number of
edges, and $f$ -- the number of triangles in the triangulation; $\chi(D)$ is 
the topological invariant of $D$ called its Euler characteristic.

The Gauss-Bonnet formula (\ref{GBtheorem}) is indeed a long way from the 
180 degree theorem, but the potential for generalization is still not 
exhausted. The ideas of Gauss about the curvature and the geometry on the 
surface was further generalized by B.~Riemann. It was soon realized that
most properties of the Riemannian geometry follows from its Levi-Civita 
parallelism, an infinitesimal parallel transport of the tangent vectors. 
The important concept of the Levi-Civita connection emerged. All these is
the mathematical basis of Einstein's general relativity. Further 
generalization of the concepts of Levi-Civita connection and curvature 
to more general, than Riemannian, manifolds lead to the notion
of fiber bundles -- the mathematical basis of the gauge field theories
\cite{1:9}. Even magnetic monopoles are related to the generalized 
Gauss-Bonnet theorem \cite{1:10}.

Therefore, both general relativity and gauge theory can be considered
as stunning generalizations of the 180 degree theorem of the Euclidean 
geometry! However, returning to the Standard Model, this is not the whole
story.
Gauge symmetry is important, very important, in the Standard Model. But the 
real shape of the world is determined by its spontaneous violation. Then a
big question is why and how the $SU(2)\times U(1)$ gauge symmetry of the 
Standard Model is broken. So far the phenomenologically adequate answer to  
this question is given by the introduction of the $SU(2)$-doublet of scalar
fields, the Higgs doublet, whose couplings and vacuum expectation value
determine fermion masses and mixings. However there are too many free 
parameters,
not fixed by the theory, indicating that in fact we do not understand what
is going on. That is why the discovery of the Higgs boson and investigation of
its properties are considered as having the paramount importance.

At this point photon colliders enter the game, because in the $\gamma\gamma$
collisions the Higgs boson will be produced as a single resonance. The idea of
photon colliders was proposed many years ago in Novosibirsk \cite{1:11,1:12}.
You have to have a linear $e^+e^-$ collider and a powerful laser (several 
Joules per flash) to realize this idea. High-energy photons are produced
by Compton backscattering of the laser light on the high-energy electrons near
the interaction point. After the scattering, the photons will have almost the
same energy as the initial electrons and small additional angular spread of 
the order of inverse $\gamma$-factor of the initial electron. This additional 
angular spread does not effect much the resulting $\gamma\gamma$ 
(or $\gamma e$) 
luminosity if the conversion point is close enough to the interaction point.
The $\gamma\gamma$ luminosity can be made even larger than the $e^+e^-$ 
luminosity at the same collider by using the initial electron beams with
smaller emittances than allowed in the $e^+e^-$-mode by beam collision 
effects.

The detailed development of the photon collider idea \cite{1:13,1:14,1:15}
showed that their construction is a quite realistic task and 
requires a small additional ($\sim 10\%$) investment compared to the linear
collider price. The solid state laser technologies with required pulse power
and duration already exist. A free electron laser with variable wave length 
is also an attractive alternative \cite{1:KPS}. 

The expected physics at high-energy photon colliders is really exciting and
very rich. It includes \cite{1:15,1:16,1:17}:
\begin{itemize}
\item Higgs boson physics, both Standard Model and supersymmetric. Especially
one should mention the unique opportunity to measure its two photon width,
as well as the possibility to explore $CP$ properties of the neutral Higgs 
boson by controlling the polarizations of the back-scattered photons.
\item Search for supersymmetry. In particular, charged sfermions, charginos 
and charged Higgs bosons will be produced at larger rates in $\gamma\gamma$
 collisions than in $e^+e^-$ collisions. The $\gamma e$ option will enable 
potential discovery of selectrons and neutralinos. The photon collider will 
also be an ideal place to discover and study stoponium bound states.
\item Exploration of the gauge bosons nonlinear interactions.
\item Top quark physics.
\item QCD-probes in a new unexplored regime.
\item Investigation of the photon structure -- its hadronic quantum 
fluctuations cannot
be completely determined from the first principles because the large 
distance effects contribute significantly. Therefore various phenomenological
models need experimental input for refinements.
\item Search for the low-scale quantum gravity, space-time noncommutativity
\cite {1:NCG} and extra dimensions.
\end{itemize}
The last item is exotic enough but one should not forget that \cite{1:18}
``Every time we introduce a new tool, it always leads to new and unexpected
discoveries, because Nature's imagination is richer than ours''.

In this paper we would like to indicate that the physical program of the 
photon collider can further be enriched if it is considered as a tool to 
perturb the QCD vacuum. An interesting phenomenon of the Disoriented Chiral
Condensate formation was discussed earlier in the context of hadron-hadron
and heavy ion collisions. We believe that photon colliders are also eligible
devices to perform such kind of research.

The paper is organized as follows. We begin with the discussion of the linear
sigma model, which is used as a QCD substitute in the majority of DCC studies. 
The idea of the Disoriented Chiral Condensate is explained and investigated
in the third section. The fourth section considers the possibility of the DCC 
production at photon colliders. The Baked Alaska scenario is examined in some
details. Quantum state of DCC is explored in the next section. It is mentioned
that at photon colliders a direct production of this state might be possible.
In the last section we provide some concluding remarks. The references on the
subject are very numerous and we list only a few of them. We hope that an
interested reader can find independently other important contributions which 
missed our attention.

\section{Linear sigma model}
The Lagrangian of quantum chromodynamics (QCD) looks ``deceptively simple''
\cite{2:1}. Indeed, it encodes the description of a surprisingly wide range of
natural phenomena, from nuclear physics to cosmology, and nevertheless is
given by the very compact expression
\begin{equation}
{\cal L}_{QCD}=\bar q (i\hat D -m)q-\frac{1}{2}Sp\, G_{\mu\nu}G^{\mu\nu},
\label{Lqcd} \end{equation}
where
$$\hat D=\hat\partial +ig\hat A,\;\; G_{\mu\nu}=\partial_\mu A_\nu-
\partial_\nu A_\mu-g[A_\mu,A_\nu],\;\; A_\mu=A_\mu^a\frac{\lambda_a}{2}$$
and $\lambda_a,\; a=1,\ldots,8$ are $SU(3)$ Gell-Mann matrices. The theory
(QCD) which is defined by this Lagrangian ``embodies deep and beautiful
principles'' and is one of ``our most perfect physical theories'' \cite{2:2}.
However, if you are interested in applying this ``most perfect physical
theory'' to understand the low-energy experimental data, you will not be
particularly happy by discovering at least three reasons \cite{2:1} for your
grievance:
\begin{itemize}
\item The Lagrangian (\ref{Lqcd}) describes quark and gluon degrees of freedom,
while ``correct'' degrees of freedom for low energy phenomena are their bound
states -- various colorless hadrons.
\item unlike quantum electrodynamics, gluons have self-interactions which
render QCD in a nonlinear theory with the corresponding increase in the
computational complexity.
\item at low energies the effective coupling constant is large and usual
perturbative methods are not applicable.
\end{itemize}
However, things are not so bad as they
look. It turns out that many important features of the low-energy dynamics are
governed by symmetries of the QCD Lagrangian and their breaking patterns. For
light quark flavors the QCD Lagrangian possesses (approximate) $U_R(3)\times
U_L(3)$ chiral symmetry. The corresponding transformations are
$$q_R\to e^{i\frac{\lambda_0}{2}\,\theta^0_R}q_R, \;\;\;
q_L\to e^{i\frac{\lambda_0}{2}\,\theta^0_L}q_L, $$
\begin{equation}
q_R\to e^{i\frac{\lambda_a}{2}\,\theta^a_R}q_R, \;\;\;
q_L\to e^{i\frac{\lambda_a}{2}\,\theta^a_L}q_L,
\label{URUL} \end{equation}
where $\lambda_0=\sqrt{\frac{2}{3}}$.
Fates of these symmetries are different.
The first line corresponds to the $U_V(1)\times U_A(1)$ transformations with
$\theta_{V,A}=\frac{1}{2}\left ( \theta^0_L\pm \theta^0_R\right )$. The
singlet vector current, generated by $U_V(1)$ transformations, remains
conserved in the low-energy limit and the corresponding conserved charge is
identified with the baryon number. On the contrary, $U_A(1)$ symmetry is broken
due to quantum anomaly. As a result, $\eta^\prime$ meson becomes much heavier
compared to other pseudoscalars. Non-Abelian symmetries
$SU_R(3)\times SU_L(3)$, as well as $U_A(1)$, are further broken spontaneously
due to a nonvanishing expectation value of the quark-antiquark condensate:
$<{\bar q}_Rq_L>\ne 0$. Eight pseudoscalar mesons ($\pi, K, \eta$) are
Goldstone bosons
associated with this symmetry breaking pattern $SU_R(3)\times SU_L(3)\to
SU_V(3)$. In fact these Goldstone bosons acquire small masses because quark
mass terms in the QCD Lagrangian break explicitly  the $U_R(3)\times U_L(3)$
chiral symmetry.

Having in mind this picture of QCD symmetries and their breaking, one can try
to model it by some effective low-energy theory for mesons, which are
excitations on the quark-antiquark condensate ground state \cite{2:3,2:3a}.
One has two kinds
of excitations, scalar and pseudoscalar mesons, because
$${\bar q}_Rq_L\sim {\bar q}q+{\bar q}\gamma_5 q.$$
Therefore, for three light quark flavors, one needs a complex $3\times 3$
matrix field $\Phi_{ab}\sim {\bar q}_{Rb}q_{La}$ to parametrize the scalar
($S$) and pseudoscalar ($P$) meson nonets:
\begin{equation}
\Phi=S+iP\equiv \frac{\lambda_a}{2}\left ( \sigma_a+i\pi_a\right )+
\frac{\lambda_0}{2}\left ( \sigma_0+i\pi_0\right ).
\label{Phi} \end{equation}
The imaginary unit is introduced to make the pseudoscalar matrix $P$
Hermitian: $iP$ corresponds to ${\bar q}\gamma_5 q$, but $({\bar q}\gamma_5 q
)^+=-{\bar q}\gamma_5 q$.

The effective Lagrangian for the field $\Phi$ should have the form \cite{2:4}
\begin{equation}
{\cal L}=Sp\,(\partial_\mu\Phi^+\partial^\mu\Phi)-V(\Phi,\Phi^+)+
{\cal L}_{SB},
\label{LSM1}\end{equation}
where ${\cal L}_{SB}$ describes symmetry breaking effects and $V(\Phi,\Phi^+)$
stands for self-interactions of the meson field. If we want the theory to be
renormalizable (although for effective theories this requirement is not
obvious), quartic couplings are at most allowed in $V(\Phi,\Phi^+)$. The
chiral transformations (\ref{URUL}) read in terms of the $\Phi$ field
\begin{eqnarray}
U_V(1) &:& \;\;\;\; \Phi\to e^{i\frac{\lambda_0}{2}\,\theta^0_V}\Phi
e^{-i\frac{\lambda_0}{2}\,\theta^0_V}=\Phi, \nonumber \\
U_A(1) &:& \;\;\;\; \Phi\to e^{i\frac{\lambda_0}{2}\,\theta^0_A}\Phi
e^{i\frac{\lambda_0}{2}\,\theta^0_A}=e^{i\lambda_0\,\theta^0_A}
\Phi, \label{ChTPhi} \\
SU_V(3) &:& \;\;\;\; \Phi\to e^{i\frac{\lambda_a}{2}\,\theta^a_V}\Phi
e^{-i\frac{\lambda_a}{2}\,\theta^a_V}, \nonumber \\
SU_A(3) &:& \;\;\;\; \Phi\to e^{i\frac{\lambda_a}{2}\,\theta^a_A}\Phi
e^{i\frac{\lambda_a}{2}\,\theta^a_A}. \nonumber
\end{eqnarray}
Therefore, $Sp\,(\Phi^+ \Phi)$ and $Sp\,(\Phi^+ \Phi)^2$ are invariant under
these transformations and the most general form of $V(\Phi,\Phi^+)$ is
\begin{equation}
V(\Phi,\Phi^+)=m^2Sp\,(\Phi^+ \Phi)+\lambda \,Sp\,(\Phi^+\Phi)^2
+\lambda^\prime \left [Sp\,(\Phi^+\Phi)\right ]^2.
\label{LSM2}\end{equation}
The symmetry breaking part of the effective Lagrangian has the form
\begin{equation}
{\cal L}_{SB}=Sp\,H(\Phi+\Phi^+)+c\left [Det(\Phi)+Det(\Phi^+)\right ].
\label{LSM3}\end{equation}
Here the first term describes explicit symmetry breaking due to nonzero
quark masses. The matrix $H$ represents the constant nine-component external
field: $H=\frac{\lambda_a}{2}h_a+\frac{\lambda_0}{2}h_0$. In practice isospin
symmetry and PCAC are good approximations because $u$ and $d$ quark
masses are very small. To preserve these symmetries, the most general
possibility is to have only two nonzero constants $h_0$ and $h_8$ \cite{2:5}.
$h_0$ gives a common shift to pseudoscalar (and scalar) masses, while $h_8$
breaks the $SU_V(3)$ unitary symmetry down to isospin $SU_V(2)$ and generates 
the mass differences between $\pi$, $K$ and $\eta$, as well as between 
their parity
partners (the phenomenological situation in the scalar nonet is not completely
clear yet \cite{2:4}). The determinant term is invariant under $SU_V(3)\times
SU_A(3)$ transformations from (\ref{ChTPhi}), because $Det(AB)=Det(A) Det(B)$
and $Det(e^{i\frac{\lambda_a}{2}\theta^a})=1$. However it violates $U_A(1)$
symmetry down to $Z_A(3)$, because $Det(e^{i\lambda_0\theta^0_A})=1$ only then
$\lambda_0\theta^0_A=\frac{2\pi}{3}n,\;n$ being an integer. This explicit
breaking of $U_A(1)$ removes the mass degeneracy between $\eta^\prime$ and
$\pi$ \cite{2:6,2:7} and, therefore, is very important for describing the
pseudoscalar nonet. Another interesting property of the determinant term is
that it gives equal and opposite sign contributions to the masses of the
corresponding scalars and pseudoscalars \cite{2:7}. Therefore, the large
splitting between scalars and pseudoscalars is expected solely from the fact
that $\eta^\prime$ is much heavier than $\pi$ \cite{2:7}. This is exactly
the situation observed in experiment. Physics behind the determinant term is
related to the $U_A(1)$ quantum anomaly, mentioned above, caused by
nonperturbative effects
in the QCD vacuum due to instantons \cite{2:8}. Note that the $i\left 
[Det(\Phi)-Det(\Phi^+)\right ]$ term is not allowed as it violates P and CP 
\cite{2:9}.
Indeed, under charge conjugation $\Phi\to\Phi^T$, which does not change the
determinant. While under parity $\Phi\to\Phi^+$ and $Det(\Phi)-Det(\Phi^+)$
changes the sign.

The linear sigma model, as defined by (\ref{LSM1}), (\ref{LSM2}) and
(\ref{LSM3}), has six free parameters to be fixed from experiment: $m^2,\;
\lambda,\;\lambda^\prime,\;c,\;h_0$ and $h_8$. Five parameters can be fixed
by using experimental information from the pseudoscalar sector alone, for
example \cite{2:4,2:10}, pion and kaon masses, the average squared mass of the
$\eta$ and $\eta^\prime$ mesons $0.5\,(m_\eta^2+m_{\eta^\prime}^2)$, and two 
decay
constants $f_\pi$ and $f_K$. To fix the $\lambda^\prime$ coupling constant, 
which violates the OZI rule \cite{2:11}, some experimental information from 
the scalar sector is required, for example \cite{2:4}, the sigma meson mass. 
The other scalar masses, the scalar and pseudoscalar mixing angles, and the
difference $m_{\eta^\prime}^2-m_\eta^2$ are then predicted quite reasonably
\cite{2:4,2:10,2:11}.

To summarize, the linear sigma model is an attractive effective theory 
candidate for description of the low energy QCD dynamics. Phenomenologically,
it is quite
successful and explains various puzzles concerning scalar and pseudoscalar
mesons \cite{2:11}:
\begin{itemize}
\item why the pion and kaon are light
\item why the $\eta^\prime$ is so heavy
\item why the scalar mesons are much heavier than pseudoscalars
\item why the sigma meson is so light compared to other scalars
\item the pseudoscalar and scalar mixing angles
\item the accidental degeneracy of the $a_0(980)$ and $f_0(980)$ mesons
\item the strong coupling of $f_0(980)$ to $K\bar K$
\item two photon widths of $a_0(980)$ and $f_0(980)$ mesons
\end{itemize}

In the next sections we will be interested in some qualitative features of
the dynamics described by the linear sigma model. At that we will make further
simplification by neglecting the effects of the strange quark. In the
two flavor case, one can assume that the field $\Phi$ in the Lagrangian is
the $2\times 2$ complex  matrix. However, $SU(2)$ has a unique property among
$SU(N)$ groups,
its fundamental representation being equivalent to its complex conjugate.
Owing to this property, two linear combinations $\Phi+\tau_2\Phi^*\tau_2$
and $\Phi-\tau_2\Phi^*\tau_2$ both transform irreducibly under the 
$SU_R(2)\times SU_L(2)$ group \cite{2:3a}. Each of them has only two 
independent complex
matrix elements. Therefore, it is possible to construct two flavor linear
sigma models by using only four lightest mass eigenstates $\pi^\pm,\,\pi^0$
and $\sigma$. Hence we take
$$\Phi=\frac{1}{2}\sigma+\frac{i}{2}\,\vec{\pi}\cdot\vec{\tau},$$
$\tau_i$ being the Pauli matrices. Then
$$\Phi^+\Phi=\frac{1}{4}\left (\sigma^2+\vec{\pi}^{\,2}\right ),\;\;
Sp\,(\Phi^+\Phi)^2=\frac{1}{2}\left [Sp\,(\Phi^+\Phi)\right ]^2=
\frac{1}{8}\left (\sigma^2+\vec{\pi}^{\,2}\right )^2$$
and, therefore, the Lagrangian takes the form (up to the irrelevant 
constant term)
\begin{equation}
{\cal L}=\frac{1}{2}\,\partial_\mu\sigma \,\partial^\mu\sigma+
\frac{1}{2}\,\partial_\mu\vec{\pi}\cdot \partial^\mu\vec{\pi}-\frac{\lambda_s}
{4}\left (\sigma^2+\vec{\pi}^{\,2}-v^2\right )^2+H\sigma,
\label{GML} \end{equation}
here
$$\lambda_s=\lambda^\prime+\frac{1}{2}\,\lambda,\;\;
v^2=-\frac{m^2}{\lambda_s} ,\;\;H=h_0.$$
This is the classic linear sigma model of Gell-Mann and Levy \cite{2:12}.
Its free parameters $\lambda_s,\;v$ and $H$ (the strength of the symmetry
preserving term, the location of its minimum and the strength of the
symmetry-breaking term) can be fixed by using pion and sigma masses and
PCAC as follows \cite{2:13}. In the chiral limit (then $H=0$) the linear
sigma model potential
$$V_S=\frac{\lambda_s}{4}\left (\sigma^2+\vec{\pi}^{\,2}-v^2\right )^2$$
has a famous ``Mexican hat'' shape. Therefore,the chiral symmetry is
spontaneously broken because the sigma field develops a nonzero vacuum
expectation value $<\sigma>=v$ (the pion field, being pseudoscalar, cannot
acquire a nonzero vacuum expectation value without violating parity). The
symmetry breaking term $V_{SB}=-H\sigma$ tilts the Mexican hat and now
$<\sigma>=\sigma_0\neq v$. Shifting the sigma field by its vacuum expectation
value, $\sigma=\sigma_0+\sigma^\prime$, and isolating quadratic terms $\frac
{m_\sigma^2}{2}\sigma^{\prime\,2}$ and $\frac{m_\pi^2}{2}\,\vec{\pi}^2$ in
the potential $V_S+V_{SB}$, we get meson masses
\begin{equation}
m_\pi^2=\lambda_s(\sigma_0^2-v^2),\;\;\;
m_\sigma^2=\lambda_s(3\sigma_0^2-v^2).
\label{mesmas} \end{equation}
The vacuum expectation value $\sigma_0$ is determined from the condition
$$\left . \frac{\partial V(\sigma,\vec{\pi})}{\partial\sigma}\right |_
{\vec{\pi}=0}=0,$$
which gives
\begin{equation}
H=\lambda_s\sigma_0(\sigma_0^2-v^2)=\sigma_0 m_\pi^2.
\label{sigma0} \end{equation}
Besides (\ref{mesmas}) and (\ref{sigma0}), we need one more relation to
determine four quantities $\lambda_s,\;v,\;H$ and $\sigma_0$. This relation
is given by PCAC:
\begin{equation}
\partial^\mu \vec{J}_{5\mu}=f_\pi m_\pi^2\,\vec{\pi}.
\label{PCAC}\end{equation}
Indeed, the axial-vector current $\vec{J}_{5\mu}$ is nothing but the Noether
current associated with the $SU_A(2)$ transformations
\begin{equation}
\Phi\to e^{i\frac{\tau_i}{2}\theta_i}\,\Phi\, e^{i\frac{\tau_i}{2}\theta_i}.
\label{su2a} \end{equation}
In terms of the $\sigma$ and $\vec{\pi}$ fields, the infinitesimal form of
Eq.\ref{su2a} reads
\begin{equation}
\delta\sigma=-\pi_i\theta_i,\;\;\; \delta\pi_i=\sigma\theta_i.
\label{su2ain} \end{equation}
The divergence of $\vec{J}_{5\mu}$ is given by the Gell-Mann-Levy equation
\cite{2:6}
$$\partial^\mu J^i_{5\mu}(x)=-\frac{\partial(\delta{\cal L})}
{\partial \theta_i(x)},$$
where $\delta{\cal L}$ is the variation of the Lagrangian under (\ref{su2ain})
with space-time dependent parameters $\theta_i(x)$, which equals
$$\delta{\cal L}=\sigma\partial_\mu\pi_i\,\partial^\mu\theta_i-
\pi_i\partial_\mu\sigma\,\partial^\mu\theta_i-H\pi_i\theta_i.$$
Therefore,
$$\partial^\mu J^i_{5\mu}(x)=-\frac{\partial(\delta{\cal L})}
{\partial \theta_i(x)}=H\pi_i$$
and comparing with PCAC Eq.\ref{PCAC} we get
\begin{equation}
H=f_\pi m_\pi^2.
\label{Hpcac} \end{equation}
Now from (\ref{mesmas}), (\ref{sigma0}) and (\ref{Hpcac}) it is easy to get
\begin{equation}
\sigma_0=f_\pi,\;\;\; \lambda_s=\frac{m_\sigma^2-m_\pi^2}{2f_\pi^2},\;\;\;
v^2=\frac{m_\sigma^2-3m_\pi^2}{m_\sigma^2-m_\pi^2}\,f_\pi^2,\;\;\; H=
f_\pi m_\pi^2.
\label{lsmpar} \end{equation}
The precise values of these parameters are largely immaterial having in mind
idealized nature of the model. In any case, they can be estimated from
(\ref{lsmpar}) if needed. For example, for $m_\sigma=600~MeV$ one gets:
$\lambda_s\sim 20,\; v\sim 90~MeV$ and $H\sim (120~MeV)^3$.

\section{Disoriented chiral condensate}
The linear sigma model potential in the limit $H\to 0$ has a degenerate
minimum at $\sigma^2+\vec{\pi}^2=v^2$ (in this limit $m_\pi=0$ and
$v=f_\pi$). The vacuum state, we believe our world is based on, points in the
$\sigma$-direction, $<\sigma>=f_\pi,\; <\vec{\pi}>=0$, and, therefore,
spontaneously violates the chiral symmetry. The natural question is whether
one can change the vacuum state by some perturbation. The following analogy
is helpful here: $SU(2)\times SU(2)$ is locally isomorphic to $O(4)$;
therefore, the order
parameter $<\Phi>$ of the linear sigma model can be considered as some analog
of spontaneous magnetization of the $O(4)$ Heisenberg ferromagnetic. Then, 
changing
the vacuum state in the relativistic field theory, which assumes an infinite
system, is analogous to rotating all spins in the infinite magnet
simultaneously and is clearly impossible. Our universe, although not infinite,
is quite large and hence at first sight we have no means to alter its vacuum
state: only one QCD vacuum state is realized in our world, all other chirally
equivalent vacuum states being unreachable and thus unphysical. However 
experience
with real magnets suggests that this simple argument (as well as virtually all
other no-go theorems) may point not so much to the real impossibility but to
the need of more elaborate imagination. In the case of ferromagnet it is
relatively simple to change the magnetization in some large enough volume. All
what is needed is to apply an external magnetic field. Even such a
comparatively weak field as Earth's magnetic field can do the job. We are
tempting here to indicate one interesting application of this effect
\cite{3:1}. Above
the Curie temperature the rotational invariance is restored in the ferromagnet
and there is no spontaneous magnetization -- all record of the previous
magnetization is lost. As lava from a volcano cools below the Curie temperature
the Earth's magnetic field aligns the magnetization of the ferromagnetic
grains. By studying such solidified lavas (basalt rocks), geophysicists have
reconstructed a history of the Earth's magnetic field with a striking result
that the Earth's magnetic field has flip-flopped many times, once in every
half million years, on the average. But this is not the most interesting part
of the story. Investigation of the ocean floor magnetization revealed a
surprising strip structure. Successive strips of normally and reversely
magnetized rock lied symmetrically on both sides of the volcanic mid-Atlantic
ridge. The explanation of this enigma comes from plate tectonics. On each
side of the ridge the tectonic plates are pulled away,
one of it towards Europa
and the other towards America. Lava, emerges from the middle,
solidifies,
sticks to the plates and is also pulled away with the magnetic field
orientation recorded in it. So the oceanic floor seems to be a gigantic tape
recorder for reversals of the Earth's magnetic field! This discovery was
crucial in recognition of Alfred Wegener's theory of  continental drift --
the idea which initially was met with enormous resistance from geophysicists.

Long ago Lee and Wick argued \cite{3:2} that an analogous domain formation
phenomenon is also possible in the case of quantum field theory with degenerate
vacuum and in principle there should exist a possibility of flipping the 
ordinary
vacuum in a limited domain of space to an abnormal one. ``The experimental
method to alter the properties of the vacuum may be called vacuum
engineering'' \cite{3:3}. It seems that a new generation of the very high 
energy
heavy ion and hadron colliders may provide a practical tool for such vacuum
engineering. The scientific significance of this possibility can hardly be
overestimated, because ``if indeed we are able to alter the vacuum, then we
may encounter some new phenomena, totally unexpected'' \cite{3:3}.

Disoriented chiral condensate formation is one of the new phenomena which may
happen in very high energy collisions \cite{3:4,3:5}. In such a collision
there is some probability that a high multiplicity final state will be produced
with high entropy. Collision debris form a hot shell expanding in all
directions nearly at the velocity of light. This shell effectively shields the
inner region up to hadronization time and then it breaks up into individual
hadrons. The hadronization time can be quite large \cite{3:6} and
during all this time the inner region has no idea about the chiral
orientation of the normal, outside vacuum. Therefore, if the inner vacuum is
perturbed enough in first instants of the collision to forget its orientation,
then almost certainly it will relax back in the ground state other than
the $\sigma$-direction. Of course, the explicit symmetry breaking ($\sim H$)
term lifts the
vacuum degeneracy. However, the corresponding tilting of the ``Mexican hat'' is
small and will not effect the initial stage evolution significantly
\cite{3:7}. Therefore, it is not unlikely that some high energy collisions can
lead to the formation of relatively large space domains where the chiral
condensate is temporarily disoriented. At later times such Disoriented Chiral
Condensate will relax back to the normal vacuum by emitting coherent
burst of pion radiation.

But how can the initial vacuum be excited? A short time after the
collision of the order of $0.3-0.8~{\mathrm fm/c}$ the energy density in the
interior of the collision region drops enough to make meaningful the 
introduction
of $\sigma$ and $\pi$ collective modes \cite{3:8}. After this time the
classical dynamics of the system is reasonably well described by the linear
sigma model. However, initially the $\sigma$ and $\pi$ fields are surrounded by
a thermal bath. So we need the sigma model at finite temperature. To reveal a
simple physical picture behind the phenomenon, we will use the following
simplified approach \cite{3:9,3:10}. Let us decompose fields into the slowly
varying classical part (the condensate) and high frequency thermal
fluctuations
$$\phi(x)=\phi_{cl}(x)+\delta\phi(x).$$
By definition the thermal average $<\phi>_{th}=\phi_{cl}$ and $<\delta
\phi>_{th}=0$. Therefore the thermal averaged symmetric potential, which
determines evolution of $\phi_{cl}$ at initial times, until the effects of the
explicit symmetry breaking term become significant, has the form (we have
suppressed isospin indices for a moment)
$$<V_S>_{th}=\left . \left . \frac{\lambda_s}{4}\right ( \phi_{cl}^2+
<(\delta\phi)^2>_{th}- v^2\right )^2.$$
To calculate $<(\delta\phi)^2>_{th}$, let us decompose $\delta\phi(x)$ into
the annihilation and creation operators
\begin{equation}
\delta\phi(x)=\int \frac{d\vec{k}}{(2\pi)^{3/2}}\frac{1}{\sqrt{2\omega_k}}
\left (a(\vec{k})e^{-ik\cdot x}+a^+(\vec{k})e^{ik\cdot x}\right ),
\label{dltphi} \end{equation}
with $\omega_k=\sqrt{\vec{k}^2+m^2}$ and (our normalization corresponds
to $(2\pi)^{-3}$ particles per unit volume)
$$[a(\vec{k}),a^+(\vec{k^{\,\prime}})]=\delta(\vec{k}-\vec{k^{\,\prime}}).$$
At the thermal equilibrium the thermal bath is homogeneous over the (large) 
spatial volume $V$. Therefore,the thermal fluctuations are the same at every
point inside $V$ and $<(\delta\phi)^2>_{th}$ can be replaced by its spatial 
average $$<(\delta\phi)^2>_{th}\to\frac{1}{V}\int d\vec{x}
<(\delta\phi)^2>_{th}.$$
Substituting here (\ref{dltphi}) we get
\begin{equation}
<(\delta\phi)^2>_{th}\to\frac{1}{V}\int \frac{d\vec{k}}{2\omega_k}<a(\vec{k})
a^+(\vec{k})+a^+(\vec{k})a(\vec{k})>_{th}.
\label{dlt2a}\end{equation}
We assumed that the chemical potential of the field $\phi$ is small, so
that the probability of finding its two quanta simultaneously in a unit
volume is negligible, and hence
$$<a(\vec{k})a(-\vec{k})>_{th}\approx 0,\;\;
<a^+(\vec{k})a^+(-\vec{k})>_{th}\approx 0. $$
However,
$$<aa^++a^+a>_{th}=2<a^+a>_{th}+[a,a^+]$$
and the second term gives a temperature independent constant. Actually this
contribution in (\ref{dlt2a}) is infinite and should be cured by
renormalization (that is subtracted). The nontrivial finite part is
$$<(\delta\phi)^2>_{th}=\frac{1}{V}\int \frac{d\vec{k}}{\omega_k}
<a^+(\vec{k})a(\vec{k})>_{th}.$$
However, $a^+(\vec{k})a(\vec{k})$ is the number density operator (in momentum
space). Hence, its thermal average is given by the Bose-Einstein
distribution
$$<a^+(\vec{k})a(\vec{k})>_{th}=\frac{N}{e^{\omega_k/T}-1},$$
where $N=V/(2\pi)^3$ is the total number of $\phi$-particles in the volume
$V$. Finally,
\begin{equation}
<(\delta\phi)^2>_{th}=\int \frac{d\vec{k}}{(2\pi)^3}~\frac{1}{\omega_k
(e^{\omega_k/T}-1)}.
\label{dlt2th} \end{equation}
In a high temperature limit $T\gg m$, (\ref{dlt2th}) is simplified to
\begin{equation}
<(\delta\phi)^2>_{th}=\int \frac{d\vec{k}}{(2\pi)^3}~\frac{1}{|\vec{k}|
(e^{|\vec{k}|/T}-1)}=\frac{T^2}{2\pi^2}\int\limits_0^\infty \frac{x\,dx}
{e^x-1}=\frac{T^2}{12}.
\label{emcontr} \end{equation}
Let us now restore the isotopic content of our theory. Each isotopic mode
gives a contribution (\ref{dlt2th}) to the effective thermal potential. However
$\sigma$-meson is too heavy. Therefore, we assume $\omega_k/T\gg 1$ for it
and neglect its contribution. There remain three pionic modes. Pions, on the
contrary, are light and we neglect their masses. Then the thermal effective
potential takes the form
\begin{equation}
<V_S>_{th}=\frac{\lambda_s}{4}\left (\sigma^2+\vec{\pi}^2+\frac{T^2}{4}-v^2
\right )^2.
\label{thpot} \end{equation}
The minimum energy configuration corresponds to
$$<\sigma>=\sqrt{v^2-\frac{T^2}{4}}.$$
Therefore, the $\sigma$-condensate completely melts down at $T_c=2v\approx
180~{\mathrm MeV}$. Above this phase transition point the vacuum configuration
corresponds to $<\sigma>\approx 0$. In fact, the $\sigma$-condensate never 
melts
completely down (for temperatures for which the linear sigma model still makes
sense), because of the $\sim H$ term. However, near the critical temperature 
this residual value of the $\sigma$-condensate (which minimizes $V\approx
\frac{\lambda_s}{4}\sigma^4-H\sigma$) is quite small
$$<\sigma>\approx \left (\frac{4H}{\lambda_s}\right )^{1/3}
\approx 3~\mathrm{MeV}\ll f_\pi.$$

Temperatures of the order $T_c$ can be reached in very high energy collisions.
Then, in some small volume, chiral condensate is melted and all information
about the ``correct'' orientation of the chiral order parameter is lost. What
happens when this volume cools down? Again an analogy with magnets is helpful.
If a magnet is heated above the Curie temperature and then slowly cooled, it
loses its spontaneous magnetization. This happens because many small domains
are formed with magnetization direction changing at random from domain to
domain, so that there is no net magnetization. Therefore, if we want to have
a large DCC domain, slow cooling in thermal equilibrium is not the best choice.
Indeed, it was argued \cite{3:11} that in such circumstances the size of DCC
domains remains small. Hopefully, the interior of the collision fireball is
cooled very rapidly due to fireball expansion. Rajagopal and Wilczek found
\cite{2:13,3:11} that in such  an out of equilibrium process larger DCC domains
can be formed. This is analogous to the quenching technique in magnet
production from a melted alloy. The physical mechanism which operates here is
the following \cite{3:7}. After a quench, the temperature suddenly drops to
zero and,
therefore, the dynamics will be governed by the zero temperature Lagrangian.
If the cooling process is very rapid, the field configuration does not have
time to follow the sudden change in the environment. Therefore,
immediately after
the quench fields do not have vacuum expectation values. Hence, the system
finds itself in a strongly out of equilibrium situation, namely near the top of
the ``Mexican hat''. The vacuum expectation values will begin to develop while
the system is rolling down towards the valley of the symmetric potential, but
this will take some time. Meanwhile the Goldstone modes (pions) will be
tachyonic:
\begin{equation}
m_\pi^2=\lambda_s\left ( <\sigma>^2-v^2\right ) <0,
\label{tachpi} \end{equation}
if $<\sigma>$ is small. Therefore, the oscillation frequencies $\omega_k=
\sqrt{\vec{k}^2+m_\pi^2}$ will be imaginary for long enough wavelengths and
they will grow with time exponentially. The zero mode is the one which is
amplified most effectively. As a result, a large sized correlated region will
be formed with nearly a uniform field. When the fields approach the bottom
of the potential and $<\sigma>$ gets close to its zero temperature value, this
mechanism ceases to operate. Therefore, a natural question is how fast the
rolling down takes place and whether the zero mode has enough time to be
significantly amplified. To answer this question, one should consider the
evolution of the $\sigma$ and $\vec{\pi}$ fields, according to the linear sigma
model. During this evolution we have
\begin{equation}
\partial^\mu\vec{J}_\mu=0,\;\; \partial^\mu\vec{J}_{5\mu}=H\vec{\pi}.
\label{conserv} \end{equation}
Derivation of the second equation (PCAC) was given earlier. At that the
axial-vector current is
$$J^i_{5\mu}=\frac{\partial(\delta{\cal L})}
{\partial(\partial^\mu \theta_i)}=\sigma\partial_\mu\pi_i-\pi_i
\partial_\mu\sigma. $$
The conserved vector current $\vec{J}_\mu$ is the Noether current associated
with the $SU_V(2)$ transformations from (\ref{ChTPhi}) and a similar procedure
will give \cite{2:6}
$$\vec{J}_\mu=\vec{\pi}\times\partial_\mu\vec{\pi}.$$
To make the problem analytically tractable, we idealize the
initial conditions and assume that the whole collision energy  is initially
localized in the infinitesimally thin pancake to an infinite transverse extent
\cite{3:12}. Then the fields can depend only on the longitudinal coordinate
$x$. Besides, such initial conditions are invariant under the longitudinal
boosts. Therefore, in fact the fields can only depend on the proper time
$\tau=\sqrt{t^2-x^2}$. Then $\partial_\mu=\frac{x_\mu}{\tau}\frac{d}{d\tau}$
with the (Minkowskian) 2-vector $x^\mu=(t,x)$. Therefore,
$$\vec{J}_\mu=\frac{x_\mu}{\tau}\,\vec{\pi}\times\dot{\vec{\pi}},\;\;\;
\vec{J}_{5\mu}=\frac{x_\mu}{\tau}\left (\vec{\pi}\dot \sigma-
\sigma\dot{\vec{\pi}}
\right ), $$
where the dot denotes differentiation with respect to $\tau$. Using
$\partial^\mu x_\mu=2$, we get
$$\partial^\mu \vec{J}_\mu=\frac{2}{\tau}\,\vec{\pi}\times
\dot{\vec{\pi}}+\tau
\frac{d}{d\tau}\left (\frac{\vec{\pi}\times\dot{\vec{\pi}}}
{\tau}\right )=\frac{1}{\tau}\frac{d}{d\tau}\left (\tau\vec{\pi}\times\dot{
\vec{\pi}}\right )$$
and
$$\partial^\mu \vec{J}_{5\mu}=\frac{1}{\tau}\frac{d}{d\tau}\left [\tau \left(
\vec{\pi}\dot \sigma-\sigma\dot{\vec{\pi}} \right ) \right ].$$
Therefore, the conservation of the vector current and PCAC (\ref{conserv})
imply
\begin{equation}
\vec{\pi}\times\dot{\vec{\pi}}=\frac{\vec{a}}{\tau},\;\;\;
\vec{\pi}\dot \sigma-\sigma\dot{\vec{\pi}}=\frac{\vec{b}}{\tau}+\frac{H}{\tau}
\int\limits_{\tau_0}^\tau \tau^\prime\vec{\pi}(\tau^\prime)\,d\tau^\prime,
\label{abeq}\end{equation}
with $\vec{a}$ and $\vec{b}$ as integration constants. Initially, far from
the valley of the symmetric potential, the symmetry breaking term
$H\vec{\pi}$ plays an
insignificant role and can be neglected. Then (\ref{abeq}) shows that $\vec{a}
\cdot \vec{b}=0$ and the triad $\vec{a},\;\vec{b},\;\vec{c}=\vec{a}\times
\vec{b}$ forms a convenient axis for decomposition of isovectors. The first
equation of (\ref{abeq}) indicates that $\pi_a=0$ and, hence, (\ref{abeq}) is
equivalent to the system
\begin{equation}
\pi_b\dot{\pi}_c-\pi_c\dot{\pi}_b=\frac{a}{\tau},\;\;
\pi_b\dot\sigma-\sigma\dot{\pi}_b=\frac{b}{\tau},\;\;
\pi_c\dot\sigma-\sigma\dot{\pi}_c=0.
\label{eqsys} \end{equation}
Because of the last equation, the motion in the ($\pi_b,\pi_c,\sigma$)-space 
is planar
$$\frac{\pi_c}{\sigma}=k=\mathrm{const}.$$
Then, the first two equations give
$$k=\frac{a}{b}.$$
To simplify the discussion, we assume $b\gg a$. Then $\pi_c\approx 0$ and the
motion plane coincides with the $(\pi_b,\sigma)$ plane. Let us introduce the
polar coordinates in this plane
\begin{equation}
\pi_b=f\sin{\theta},\;\;\; \sigma=f\cos{\theta}.
\label{polarc} \end{equation}
Then, (\ref{eqsys}) gives
\begin{equation}
f^2\dot \theta=-\frac{b}{\tau}.
\label{eqtheta}\end{equation}
The equation for the radial coordinate $f$ can be derived from the equation
of motion
$$\Box\vec{\pi}=-\lambda_s\left (\sigma^2+\vec{\pi}^2-v^2\right )\vec{\pi}$$
with $\Box=\partial_\mu\partial^\mu$ as the d'Alembertian.
In our case, this equation is equivalent to
$$\frac{1}{\tau}\,\frac{d}{d\tau}\left (\tau\frac{d\pi_b}{d\tau}\right )=
-\lambda_s\left (f^2-v^2\right )\pi_b.$$
Substituting here (\ref{polarc}) and using (\ref{eqtheta}), we get the
radial equation
\begin{equation}
\ddot f+\frac{\dot f}{\tau}=\frac{b^2}{f^3\tau^2}-\lambda_s
\left (f^2-v^2\right )f.
\label{eqradial} \end{equation}
At later times the difference $g=(f-v)/v$ is expected to be small. Therefore,
$$\left (f^2-v^2\right )f\approx 2v^3g$$
and (\ref{eqradial}) reduces to
$$ \ddot g+\frac{\dot g}{\tau}=\frac{b^2}{v^4\tau^2}-2\lambda_s v^2g. $$
Introducing a new dimensionless variable $s=\sqrt{2\lambda_s}v\tau$ (note that
in the $H=0$ limit $m_\sigma=\sqrt{2\lambda_s}v$), we get the
inhomogeneous Bessel equation
\begin{equation}
s^2\,\frac{d^2g}{ds^2}+s\,\frac{dg}{ds}+s^2g=\left (\frac{b}{v^2}\right )^2.
\label{Beseq} \end{equation}
Therefore, the solution  can be expressed through the Bessel functions
$J_0(s)$ and $Y_0(s)$. Hence, for large
proper times it will exhibit a damped oscillatory behavior. For example, for
large $s\gg 1$,
$$J_0(s)\approx \sqrt{\frac{2}{\pi s}}\,
\cos{\left (s-\frac{\pi}{4}\right )}.$$
Large enough compared to what number? The inhomogeneous term in (\ref{Beseq}),
which is a reminiscence of the influence of the angular motion on the radial
motion, is characterized by a dimensionless number $b/v^2$, which we assume
to be much greater than one. Therefore, the asymptotic value of $s$ can be 
estimated to be $s\sim b/v^2$, which translates into the proper time
\begin{equation}
\tau_R\sim \frac{b}{\sqrt{2\lambda_s} v^3}.
\label{tauR} \end{equation}
This gives us an estimate of the rolling-down time.

Let us now consider the process of formation and growth of correlated domains
in a scalar quantum field theory after a quench from an equilibrium disordered
initial state at the temperature $T_i=T$ to a final state at $T_f\approx 0$
\cite{3:13}. For unstable modes the instantaneous quench will be mimicked by
a time dependent mass
$$m^2(t)=m_i^2\,\Theta(-t)-m_f^2\,\Theta(t),$$
which is tachyonic at $t>0$. In the decomposition
\begin{equation}
\phi(\vec{x},t)=\int \frac{d\vec{k}}{(2\pi)^{3/2}}\frac{1}{\sqrt{2\omega_k}}
\left (a(\vec{k})u_k(t)e^{i\vec{k}\cdot \vec{x}}+
a^+(\vec{k})u^*_k(t)e^{-i\vec{k}\cdot \vec{x}}\right ),
\label{phidec} \end{equation}
the corresponding mode functions $u_k(t)$ obey
$$\left [ \frac{d^2}{dt^2}+\vec{k}^2+m^2(t)\right ] u_k(t)=0$$
and initially (for $t<0$) we have $u_k(t)=e^{-i\omega_k t},\;
\omega_k=\sqrt{m_i^2+\vec{k}^2}$. For $t>0$ the solution is
\begin{equation}
u_k(t)=A_k e^{\alpha_k t}+B_k e^{-\alpha_k t},
\label{uk}\end{equation}
with $\alpha_k=\sqrt{m_f^2-\vec{k}^2}$ (we will concentrate on unstable modes
so that $\vec{k}^2<m_f^2$). Matching the $t>0$ and $t<0$ solutions and their
first derivatives at $t=0$, we can determine the $A_k$ and $B_k$ coefficients
\begin{equation}
A_k=\frac{1}{2}\left (1-i\,\frac{\omega_k}{\alpha_k}\right ), \;\;\;
B_k=\frac{1}{2}\left (1+i\,\frac{\omega_k}{\alpha_k}\right ).
\label{AkBk}\end{equation}

The information about the domain size is encoded in the equal time correlation
function (spatially averaged over the volume $V$)
\begin{equation}
G(\vec{x},t)=\frac{1}{V}\int d\vec{y}<\phi(\vec{x}+\vec{y},t)\phi(\vec{y},t)
>_{th}.
\label{corrfun} \end{equation}
Indeed, if $x$ is not greater than the domain size $L_D$, then $\phi(\vec{x}+
\vec{y},t)\phi(\vec{y},t)\approx \phi^2(\vec{y},t)$ and the integral
(\ref{corrfun}) should be near its maximal value $G(\vec{0},t)$. On the 
contrary, if $x\gg L_D$, then the integral (\ref{corrfun}) averages to zero.
Substituting (\ref{phidec}) into (\ref{corrfun}) and remembering that for our
approximations
$$<aa>_{th}\approx 0, \;\;\; <a^+a^+>_{th}\approx 0,$$
we get
$$G(\vec{x},t)=$$
$$=\frac{1}{V}\int \frac{d\vec{k}}{2\omega_k}|u_k(t)|^2\left [
<a(\vec{k})a^+(\vec{k})>_{th}e^{i\vec{k}\cdot \vec{x}}+
<a^+(\vec{k})a(\vec{k})>_{th}e^{-i\vec{k}\cdot \vec{x}}\right ]=$$
\begin{equation}
=\frac{1}{V}\int \frac{d\vec{k}}{2\omega_k}|u_k(t)|^2\left [
<a(\vec{k})a^+(\vec{k})>_{th}+<a(-\vec{k})a^+(-\vec{k})>_{th}+
\delta(\vec{0})\right ]e^{i\vec{k}\cdot
\vec{x}}.
\label{Gxt} \end{equation}
To understand the meaning of $\delta(\vec{0})$, let us return one step
backward and write
$$\frac{|u_k(t)|^2}{2\omega_k}\,\delta(\vec{0})=\int\frac{d\vec{k^{\,\prime}}}
{(2\pi)^3}\,\delta(\vec{k}-\vec{k^{\,\prime}})\,e^{-i\vec{y}\cdot (\vec{k}-
\vec{k^{\,\prime}})}\,\frac{u^*_k(t)}{\sqrt{2\omega_k}}\,
\frac{u_{k^{\,\prime}}(t)}
{\sqrt{2\omega_{k^\prime}}}\,d\vec{y}=$$
$$=\frac{|u_k(t)|^2}{2\omega_k}\int\frac{d\vec{y}}{(2\pi)^3}=
\frac{|u_k(t)|^2}{2\omega_k}\,\frac{V}{(2\pi)^3}.$$
Besides, as was explained above,
$$<a^+(\vec{k})a(\vec{k})>_{th}=\frac{N}{e^{\omega_k/T}-1}, \;\;\;
N=V/(2\pi)^3.$$
Therefore, (\ref{Gxt}) takes the form
$$G(\vec{x},t)=\int \frac{d\vec{k}}{(2\pi)^3}\,\frac{1}{2\omega_k}\,
|u_k(t)|^2\,
\coth{\left(\frac{\omega_k}{2T}\right )}\,e^{i\vec{k}\cdot \vec{x}}.$$
After the quench $T\approx 0$ and hence $ \coth{\left(\frac{\omega_k}{2T}
\right )}\approx 1$. To study the growth of domains, one should subtract
the contributions that were already present before the quench \cite{3:13}
$$\tilde G(\vec{x},t)=G(\vec{x},t)-G(\vec{x},0)=
\int \frac{d\vec{k}}{(2\pi)^3}\,\frac{1}{2\omega_k}\,
\left [|u_k(t)|^2-1\right ]\,e^{i\vec{k}\cdot \vec{x}}.$$
But (\ref{uk}) and (\ref{AkBk}) imply
$$|u_k(t)|^2=\frac{1}{2}\left (1+\frac{\omega_k^2}{\alpha_k^2}\right )
\cosh{(2\alpha_k t)}+\frac{1}{2}\left (1-\frac{\omega_k^2}{\alpha_k^2}
\right ). $$
Therefore, the contribution of unstable modes in the domain growth is
controlled by
$$\tilde G(\vec{x},t)=\frac{1}{2}\int \frac{d\vec{k}}{(2\pi)^3}\,
\frac{1}{2\omega_k}\,\left (1+\frac{\omega_k^2}{\alpha_k^2}\right )
\left[\cosh{(2\alpha_k t)}-1\right ]\,e^{i\vec{k}\cdot \vec{x}}=$$
\begin{equation}
=\frac{1}{4\pi^2 r^2}\int\limits_0^{m_f} (kr)\sin{(kr)}
\frac{1}{\omega_k}\,\left (1+\frac{\omega_k^2}{\alpha_k^2}\right )\,
\sinh^2{(\alpha_k t)}\, dk,
\label{tildeG} \end{equation}
where $r=|\vec{x}|$ and $k=|\vec{k}|$. We are interested in the large $t$
asymptote of this function. Then
$$\sinh^2{(\alpha_k t)}\to e^{2\alpha_k t}$$
and
$$\tilde G(\vec{x},t)\to \frac{m_i^2+m_f^2}{16\pi^2 r^2}\int\limits_0^{m_f}
dk\,\frac{\sin{(kr)}}{\omega_k\alpha_k^2}\,e^{g(k)},$$
where
$$g(k)=2\alpha_k t+\ln{(kr)}.$$
Note that $g(k)$ has a maximum at $k_0\approx \sqrt{\frac{m_f}{2t}}$ and
$$g^{\prime \prime}(k_0)\approx -\frac{4t}{m_f}.$$
Therefore
$$g(k)\approx g(k_0)-\frac{2t}{m_f}\,(k-k_0)^2$$
and for large $t$ the function $e^{g(k)}$ has a very sharp peak at $k=
k_0$. Near this point $(\omega_k\alpha_k^2)^{-1}$ is a slowly varying function.
Therefore
$$\tilde G(\vec{x},t)\to \frac{k_0}{16\pi^2r}\,\frac{m_i^2+m_f^2}{m_im_f^2}
\;e^{2m_ft}\,I,\;\;\; I=\int\limits_0^{m_f}
dk\,e^{-\frac{2t}{m_f}\,(k-k_0)^2}\,\sin{(kr)},$$
where we have used
$$\frac{1}{\omega_{k_0}\alpha_{k_0}^2}\approx \frac{1}{m_i m_f^2},\;\;\;
g(k_0)\approx 2m_ft + \ln{(k_0 r)}.$$
The remaining integral 
$$I\approx \int\limits_{-\infty}^\infty dk\,e^{-\gamma\,(k-k_0)^2}\,\sin{(kr)}=
\sqrt{\frac{\pi}{\gamma}}\,\sin{(k_0r)}\,e^{-\frac{r^2}{4\gamma}},\;\;
\gamma=\frac{2t}{m_f}.$$
As we can see \cite{3:13,3:14}
$$\tilde G(\vec{x},t)\approx \tilde G(\vec{0},t)\,\frac{\sin{(k_0r)}}
{k_0r}\,e^{-\frac{m_fr^2}{8t}}.$$
Therefore, the domain size grows with time, according to the Cahn-Allen scaling
relation \cite{3:14,3:15}
\begin{equation}
L_D(t)=\sqrt{\frac{8t}{m_f}}.
\label{CASR} \end{equation}
Remembering our estimate (\ref{tauR}) for the rolling-down time and taking
a mean value of (\ref{tachpi}) as an estimate for $m_f^2$:
$m_f^2=\frac{1}{2}\lambda_s v^2,$ we get
\begin{equation}
L_D=\frac{1}{v}\sqrt{\frac{8b}{\lambda_s v^2}}\approx 1.4~\mathrm{fm} \;
\sqrt{\frac{b}{v^2}}.
\label{domsize} \end{equation}
Assuming Gaussian initial fluctuations of the fields and of their derivatives,
it can be shown \cite{3:12} that the probability of the initial strength $b$
of the axial-vector current to be large is exponentially suppressed. Therefore,
our estimate (\ref{domsize}) shows that typically DCC domains are quite small,
unless $b$ is large enough in some rare occasions. One concludes that
``formation of an observable DCC is likely to be a rather natural but rare
phenomenon'' \cite{3:16}.

To close this section, let us indicate some review articles about the DCC
phenomenon \cite{3:6,3:8,3:16,3:17}, where an interested reader can find 
further discussions and references to the original literature, which is quite
numerous.

\section{DCC at a photon collider}
Usually DCC formation is considered in the context of heavy ion or 
hadron-hadron collisions. We see no reason why gamma-gamma collisions have
to be discriminated in this respect. The basic interaction for the photon is
of course an electromagnetic interaction with charged particles, and to lowest
order in $\alpha$ the photon appears as a point-like particle in its 
interactions. However, according to quantum field theory, the photon may 
fluctuate into a virtual charged fermion-antifermion pair. In this way  
strong interactions come into play through quark-antiquark fluctuations.
While the high-virtuality part of such quark-antiquark fluctuations can be
calculated perturbatively, the low-virtuality part cannot. The latter is
usually described phenomenologically by a sum over low-mass vector-meson 
states -- the Vector Meson Dominance (VMD) ansatz. Therefore, the effective
photon state vector has the form \cite{4:1,4:2}
\begin{equation}
|\gamma>=\sqrt{Z_3}|\gamma_{bare}>+\sum\limits_V c_V|V>+\sum\limits_q c_q
|q\bar q>+\sum\limits_l c_l|l^+ l^->,
\label{gamma}\end{equation}
where \cite{4:1}
$$c_V^2=\frac{4\pi\alpha}{f_V^2},\;\;\frac{f_\rho^2}{4\pi}\approx 2.2,\;\;
\frac{f_\omega^2}{4\pi}\approx 23.6,\;\;\frac{f_\phi^2}{4\pi}\approx 18.4,
\;\;\frac{f_{J/\Psi}}{4\pi}\approx 11.5 .$$ 
The coefficients of the perturbative $|q\bar q>$ part depend on the scale 
$\mu$ at that the photon is probed. Namely \cite{4:1}
$$c_q^2=\frac{\alpha}{\pi}e_q^2\ln{\frac{\mu^2}{k_0^2}},$$
where $|e_q|=\frac{1}{3},\frac{2}{3}$ is the quark charge and $k_0$ is an
unphysical parameter separating the low- and high-virtuality parts of the
quark-antiquark fluctuations.

The last term in (\ref{gamma}) describes fluctuations into lepton pairs and
is uninteresting inasmuch as the hadronic final state is concerned. The
coefficient of the first bare-photon term is given by
$$Z_3=1-\sum\limits_V c^2_V-\sum\limits_q c^2_q-\sum\limits_l c^2_l$$
and is close to unity.

Therefore, for some fraction of time the photon behaves like a hadron. This
fraction is quite small, about $1/400$ \cite{4:3}, but for hadronic final
states this smallness is overcompensated by the fact that in its hadron facet
the photon experiences strong interactions. 

According to (\ref{gamma}), one has six different types of possible 
interactions in the high-energy photon-photon collisions \cite{4:1}:
\begin{itemize}
\item both photons turn into hadrons (vector mesons) and the partons of these
hadrons interact with each other.
\item one photon turns into a hadron and its partons interact with the
quark-antiquark fluctuation of another photon.
\item both photons fluctuate perturbatively into quark-antiquark pairs and 
subsequently these fluctuations interact with each other.
\item a bare photon interacts with the partons of the hadron which another
photon was turned into.
\item a bare photon interacts with the quark-antiquark fluctuation of
another photon. 
\item bare photons interact directly in a hard process.
\end{itemize}  
In fact, in the total hadronic cross sections, the first two event classes
dominate, the bulk of the contribution coming from the $\rho^0\rho^0$ 
component 
of the first class \cite{4:1}. Therefore, high-energy photon-photon collisions
are very much similar to the hadron-hadron collisions, and if DCC can be 
formed in the latter case, it will be formed also in the former case. Of 
course, the photon-photon cross section is strongly reduced compared to the
hadron-hadron cross sections (about $10^5$ times). However, it is not 
improbable
that a great deal of this smallness is overcome by somewhat more favorable
conditions for the DCC formation in the gamma-gamma collisions than in the
proton-(anti)proton collisions. The argument goes as follows. As has been 
mentioned
above, for boost-invariant initial conditions, when the field depends only on
the proper time $\tau=\sqrt{t^2-x^2}$, the d'Alembertian equals
$$\Box=\frac{1}{\tau}\,\frac{d}{d\tau}\left (\tau\frac{d}{d\tau}\right )
=\frac{d^2}{d\tau^2}+\frac{1}{\tau}\;\frac{d}{d\tau}.$$
The second term describes the decrease of energy in a covolume due to 
longitudinal expansion and brings an effective ``friction'', which is 
necessary for quenching, into the equation of motion \cite{3:12,4:4}.
The transverse
($D=2$) and spherical ($D=3$) expansions can be modeled analogously if one
assumes that the field depends only on  $\tau=\sqrt{t^2-\sum\limits_{i=1}^D
x_i^2}$. Then
$$\Box=\frac{d^2}{d\tau^2}+\frac{D}{\tau}\;\frac{d}{d\tau}.$$ 
Therefore, the larger is $D$ the more efficient is the quenching 
and the spherical expansion seems to be the most favorable
for pion zero mode amplification \cite{4:4,4:5}. This simple observation
is confirmed by a more detailed study \cite{4:5}. However, to organize an 
isotropically expanding fireball is not a trivial task even in head-on 
hadron-hadron collisions. Constituent quarks inside hadrons become ``black''
at high energies, and for the projectile remnants not to spoil the isotropic 
expansion,
one may wait for a rare event when these black disks inside the projectiles are
aligned \cite{4:6}. The probability that all six constituents are aligned in
colliding protons during a head-on collision is $p_1\sim \left (\frac{r_q^2}
{r_p^2}\right )^5$, while the analogous probability for the four constituents
of $\rho^0\rho^0$ collisions is $p_2\sim \left (\frac{r_q^2}{r_\rho^2}
\right )^3$. Taking $r_q\approx \frac{1}{2}r_\rho\approx \frac{1}{3}r_p$,
we get $p_2/p_1\approx 10^3$. Therefore, photon-photon collisions seem to be
more favorable in this respect. 

Even if the ``right'' fireball is prepared, the odds of the large DCC domain
formation are usually small. In \cite{4:4} this probability was found to be
about $10^{-3}$. Remember that our preceding considerations indicate that one 
needs large initial strength of the axial-vector $SU(2)$-current. In 
gamma-gamma
collisions this initial strength is expected to be enhanced due to chiral 
anomaly effects, analogous to what was considered in \cite{4:7} for heavy-ion
collisions. The effects of chiral anomaly can be incorporated in the linear 
sigma model by adding the following interaction Lagrangian
$${\cal L}_{anom.}=\frac{\alpha}{4\pi f_\pi}\,\pi^0
\epsilon_{\mu\nu\sigma\tau}F^{\mu\nu}F^{\sigma\tau}= 
\frac{\alpha}{\pi f_\pi}\,\pi^0\vec{E}\cdot\vec{H}.$$
Under $SU_A(2)$ transformations (\ref{su2ain}), we have
$$\delta{\cal L}_{anom.}=\frac{\alpha}{\pi f_\pi}\,\vec{E}\cdot\vec{H}\,
\sigma\,\theta_3.$$
Therefore, according to the Gell-Mann-Levi equation (we neglect explicit
symmetry breaking):
$$\partial^\mu J^3_{5\mu}(x)=-\frac{\partial(\delta{\cal L})}
{\partial \theta_3(x)}=-\frac{\alpha}{\pi f_\pi}\,\vec{E}\cdot\vec{H}\;
\sigma$$
and the axial-vector current is no longer conserved even in the limit of zero
quark masses. As a result, electromagnetic fields can lead to the enhancement
of the axial-vector current strength in the $\pi^0$-direction. The 
corresponding induced strength equals
\begin{equation}
b_3=-\frac{\alpha}{\pi f_\pi}\int \vec{E(\tau)}\cdot\vec{H(\tau)}\,
\sigma(\tau)\,\tau d\tau.
\label{b3anom} \end{equation}
As was shown in \cite{4:7}, the expected effects are small in relativistic
heavy-ion collisions, but nevertheless this initial small ``kick'' can have
substantial effect on the DCC formation. Unfortunately we can not use 
(\ref{b3anom}) to estimate how big is the kick in gamma-gamma collisions --
the application of the sigma model makes sense only after some time after the
collision, while the effects of the chiral anomaly on the $b_3$ magnitude
are confined to the first instants of the collision. 

It was suggested \cite{4:8,4:9} that in hadron-hadron collisions DCC could be
formed through  the ``Baked Alaska'' scenario. However, as we have mentioned 
above, the
considerable part of the $\gamma\gamma\to hadrons$ cross section
is due to the $\rho^0\rho^0$ mechanism. Therefore, the Baked Alaska model 
should work for gamma-gamma collisions too. So let us take a closer look 
at it.

Normally Baked Alaska is a delightful dessert where ice cream is covered by
meringue and then baked very quickly in a hot oven without melting the ice
cream (you can find the recipe in the
appendix. Try it and enjoy). In physics context, however, this term firstly
appeared as denoting a model for nucleation of the $B$ phase of superfluid
$^3He$ inside the supercooled $A$ phase \cite{4:10,4:11,4:12}. The surface 
tension at 
the boundary between the $A$ and $B$ phases is anomalously large. Therefore 
usual
small bubbles of the $B$ phase, created inside the $A$ phase by thermal
fluctuations, are energetically not profitable. Hence, they shrink and vanish. 
Only for a very large bubble
the volume energy gain overcomes the surface energy and the bubble begins to
grow. However, it is virtually impossible to create such a gigantic critical 
bubble
by thermal fluctuations. The experiment, nevertheless, discovered a high enough
nucleation rate. To explain the puzzle, Leggett suggested that the nucleation
was assisted by cosmic rays \cite{4:10}. Secondary electrons from the passage
of a cosmic-ray muon through the liquid create hot spots in $^3He$ by 
depositing several hundred eV energy in small volumes. Inside such a 
``fireball'' the Cooper pairs of the $^3He$ atoms are broken and, therefore, 
the
normal, Fermi-liquid phase of the $^3He$ is restored. Yet, the fireball expands
quickly and becomes a ``Baked Alaska'': a cold core surrounded by a hot, thin
shell of normal fluid. There is some probability that after the core is cooled
below the superfluidity phase transition temperature, it finds itself in the
$B$ phase. This $B$ phase core can expand to larger than the critical radius
because for some time it is shielded from the $A$ phase bulk by the expanding 
hot
shell, thereby eliminating surface energy price of the $A-B$ boundary layer.
When the shielding shell finally disappears, the $B$ phase bubble is larger
than the critical one and, therefore, expands further until it fills the whole
vessel.  

Let us return to Baked Alaskas produced by high-energy gamma-gamma collisions.
Suppose DCC is formed inside the Baked Alaska core with the misalignment angle
$\theta$. That is inside the DCC region one has
$$<\sigma>_{DCC}=f_\pi\cos{\theta},\;\;\; <\vec{\pi}>_{DCC}=f_\pi\sin{\theta}
~\vec{n},$$
$\vec{n}$ being a unit vector in isospin space. Outside the fireball one has
the normal vacuum:
$$<\sigma>=f_\pi,\;\;\;<\vec{\pi}>=0.$$
Finally, when the shielding shell of hot hadronic matter disappears, the DCC
relaxes to this outside normal vacuum by emitting coherent low energy pions.
Hadronization of the shell also produces mainly pions and, therefore, generates
a background to the DCC signal. Simple considerations allow one to estimate the
numbers of the DCC and background pions \cite{4:9}. Energy density in the DCC
region is higher than in the normal vacuum because of the symmetry breaking 
term
$V_{SB}=-H\sigma$. The difference is
$$\Delta \epsilon_V=-H<\sigma>_{DCC}+H<\sigma>=
Hf_\pi(1-\cos{\theta})=2f_\pi^2 m_\pi^2\sin^2{\frac{\theta}{2}}.$$
Therefore, the total volume energy available for pion radiation from the DCC
decay is
$$E_V=\frac{8\pi}{3}R^3f_\pi^2m_\pi^2\sin^2{\frac{\theta}{2}}\, ,$$
where $R$ is the fireball radius at the moment of hadronization. However, pions
radiated from the DCC are nonrelativistic (in the DCC rest frame). Therefore,
the expected average number of such pions is (we have assumed that 
$<\sin^2{\frac{\theta}{2}}>=\frac{1}{2}$)
\begin{equation}
N_V\approx\frac{E_V}{m_\pi}=\frac{4\pi}{3}R^3f_\pi^2\,m_\pi.
\label{NVpions} \end{equation}
One can assume \cite{4:9} that at the moment of hadronization the shell
consists of one densely packed layer of pions, each having the radius
$r_\pi\approx \frac{1}{2}m_\pi^{-1}\approx 0.7~\mathrm{fm}$. Therefore, the
number of background pions from the fireball shell is
\begin{equation}
N_b\approx \frac{4R^2}{r_\pi^2}.
\label{Nbpions} \end{equation}
For a large DCC bubble of the radius $R\approx 10 r_\pi\approx 7~\mathrm{fm}$,
the above given estimates imply
$$N_V\approx\frac{4\pi}{3}\,125\left (\frac{f_\pi}{m_\pi}\right)^2\approx
250,\;\;\; N_b\approx 400.$$
There will also be coherent pions associated with the surface energy of the 
interface between the DCC and outside vacuum. The energy density in the 
interface is dominated by the contribution due to gradients of the fields. If 
one
assumes that the interface thickness is the same as for the hadronized shell,
that is $d\approx 2r_\pi\approx m_\pi^{-1},$ then
$$\epsilon_S\approx\frac{1}{2}\left [(\Delta\vec{\pi})^2+(\Delta\sigma)^2
\right]\approx\frac{f_\pi^2}{2d^2}\left[\sin^2{\theta}+(1-\cos{\theta})^2
\right]=2\frac{f_\pi^2}{d^2}\,\sin^2{\frac{\theta}{2}}.$$
The corresponding total average energy is thus
$$E_S\approx 4\pi\,R^2\,d\,<\epsilon_S>\approx 4\pi\,R^2\,f_\pi^2\,m_\pi.$$
Pions originated from the surface layer of thickness $d$ will have 
characteristic momenta $p_\pi\sim 1/d\approx m_\pi$ and energy $E_\pi=\sqrt{
p_\pi^2+m_\pi^2}\approx\sqrt{2}m_\pi$. Therefore the expected average number
of surface pions is
$$N_S\approx \frac{E_S}{E_\pi}\approx 2\sqrt{2}\pi R^2\,f_\pi^2\approx
50\sqrt{2}\pi\left (\frac{f_\pi}{m_\pi}\right)^2\approx 105.$$ 

As we can see, the DCC signal from such a large single domain is quite 
prominent.
This becomes especially evident if one realizes that there is a large 
probability, $\sim 10\%$, that almost all of these 250 nonrelativistic 
signal pions are charged ones, with only a few neutral pion admixture. The 
probability that such a huge isospin-violating fluctuation happens in the 
background pions is, of course, completely negligible. This striking feature 
of the DCC signal follows from the following simple geometrical argument 
\cite{4:13}
(the inverse-square-root distribution, discussed below, was independently 
rediscovered many times by different authors. See \cite{3:6} for relevant 
references). Let the pion field in a single DCC domain be aligned along a 
fixed isospin direction $\vec{n}=(\sin{\theta}\cos{\phi},\, \sin{\theta}
\sin{\phi}, \, \cos{\theta})$. Classically the radiation is proportional to 
the square of the field strength. Therefore the fraction of neutral pions $f$,
emitted during relaxation of such a DCC domain, equals
\begin{equation}
f=\frac{|\pi_3|^2}{\sum\limits_{a=1}^3|\pi_a|^2}=\cos^2{\theta}.
\label{fdef} \end{equation}
The probability for $f$ to be in the interval $(f,\,f+df)$ is given by
\begin{equation}
P(f)df=\left [P(\cos{\theta})+P(-\cos{\theta})\right ]\,d\cos{\theta}.
\label{Pfdf} \end{equation}
Any orientation of the unit vector $\vec{n}$ is equally valid. Therefore the
probability $P(\cos{\theta})d\cos{\theta}$ for finding $\cos{\theta}$ in the
interval $(\cos{\theta},\,\cos{\theta}+d\cos{\theta})$ equals
$$\frac{1}{4\pi}\int\limits_{\cos{\theta}}^{\cos{\theta}+d\cos{\theta}}
d\cos{\theta}\int\limits_0^{2\pi}d\phi=\frac{1}{2}\,d\cos{\theta}.$$
This implies $P(\cos{\theta})=\frac{1}{2}$. Therefore, from (\ref{Pfdf}) the
probability density for the neutral fraction $f$ is
$$P(f)=\frac{d\cos{\theta}}{df},$$
while from (\ref{fdef}) $df=2\cos{\theta}\,d\cos{\theta}=2\sqrt{f}\,
d\cos{\theta}$, and we finally obtain 
\begin{equation}
P(f)=\frac{1}{2\sqrt{f}}.
\label{ISQRTD} \end{equation}
This inverse-square-root distribution is drastically different from what is
expected for noncoherent pion production: the binomial-distribution which for
large pion multiplicities $N$ turns into a narrow Gaussian centered at 
$f=\frac{1}{3}$:
$$P_{nc}(f)=C_N^{Nf}\left (\frac{1}{3}\right )^{Nf} \left (\frac{2}{3}\right )
^{N(1-f)}\longrightarrow \frac{3N}{2\sqrt{\pi}}\;e^{-\frac{9N\left ( 
f-\frac{1}{3}\right )^2}{4}}.$$
For example, the probability that $f$ does not exceeds $0.01$ according to
(\ref{ISQRTD}) equals
$$P(f\le 0.01)=\int\limits_0^{0.01}\frac{df}{2\sqrt{f}}=\sqrt{0.01}=
10\% \;\;!\,!\,,$$
while the binomial-distribution predicts
$$P(f\le 0.01)=\int\limits_0^{0.01} P_{nc}(f)\,df\approx 0.01\,\frac{3N}
{2\sqrt{\pi}}\;
e^{-\frac{N}{4}}\ll 1,\;\;\mathrm{for}\;N\gg 1.$$

However, as we have seen above, it is not easy to produce large DCC domain. For
many small DCC domains, with random vacuum orientations, the effect of the
inverse-square-root distribution will be washed out by averaging over the
orientations (it was, however, argued in \cite{4:14} that the later case of 
small DCC domains may lead to enhanced baryon-antibaryon production within
the framework of the Skyrmion picture of the nucleon). The intuitive reason
why it is difficult to grow up a large DCC bubble is the following. Up to now
the only mechanism discussed by us for the DCC formation, was the spinodal 
instability which operates when the fields are rolling-down from the top of
the Mexican hat potential towards the valley. But this rolling-down time is
small unless the initial strength of the axial-vector current is large (see
Eq. \ref{tauR}). Fortunately, there exists a parametric resonance mechanism
\cite{4:15} which can further assist the DCC formation, after the spinodal
instability is over. To illustrate the physical idea, let us again return to
the Blaizot-Krzywicki model \cite{3:12} with boost-invariant initial 
conditions. The equation of motion for the pion field is 
\begin{equation}
\Box\vec{\pi}=-\lambda_s\left (\sigma^2+\vec{\pi}^2-v^2\right )\vec{\pi}
\label{pieqm} \end{equation}
with 
$$\Box=\frac{d^2}{d\tau^2}+\frac{1}{\tau}\;\frac{d}{d\tau}.$$
At later times the fields are near the true vacuum. So we assume
$$\sigma\approx f_\pi,\;\; \vec{\pi}=\pi(\tau)\vec{n}.$$
The above pionic field describes a (small) disoriented chiral condensate 
aligned along a fixed unit vector $\vec{n}$ in isospace -- the result of
preceding spinodal instability. If we neglect nonlinear terms, we get in the
zeroth approximation (note that $\lambda_s\left (f_\pi^2-v^2\right )=
m_\pi^2$ )
\begin{equation}
\ddot \pi^{(0)}(\tau)+\frac{\dot\pi^{(0)}(\tau)}{\tau}+m_\pi^2\, 
\pi^{(0)}(\tau)=0.
\label{pieqm0} \end{equation}  
This equation is equivalent to the Bessel equation and its general solution
is a linear combination of the Bessel functions $J_0(m_\pi\tau)$ and 
$Y_0(m_\pi\tau)$. Therefore in the large $\tau$ limit one expects damped
oscillations
$$\pi^{(0)}(\tau)=\frac{A}{\sqrt{\tau}}\,\cos{(m_\pi\tau+\varphi)},$$
with $A$ and $\varphi$ as some constants. Now consider a fluctuation 
(in the $\vec{n}$-direction)
$\pi^{(1)}(\tau)$ around the zero-mode $\pi^{(0)}(\tau)$: $\pi(\tau)=
\pi^{(0)}(\tau)+\pi^{(1)}(\tau)$. Keeping only the terms linear in
$\pi^{(1)}(\tau)$ we get from (\ref{pieqm}) and (\ref{pieqm0})
$$\ddot \pi^{(1)}(\tau)+\frac{\dot\pi^{(1)}(\tau)}{\tau}+m_\pi^2\, 
\pi^{(1)}(\tau)=-3\lambda_s \pi^{(0)\,2}(\tau)\,\pi^{(1)}(\tau),$$
or (we have neglected unimportant $\frac{3\lambda_s A^2}{2\tau}\,\pi^{(1)}
(\tau)$ term)
$$\ddot \pi^{(1)}(\tau)+\frac{\dot\pi^{(1)}(\tau)}{\tau}+\omega_0^2\left [
1+\frac{q}{\tau}\cos{(\omega\tau+2\varphi)}\right ]\pi^{(1)}(\tau)=0,$$
where $\omega_0=m_\pi$, $\omega=2m_\pi$ and $q=\frac{3\lambda_s A^2}
{2m_\pi^2}$. Therefore, one expects a parametric resonance because $\omega=
2\omega_0$.

For more general initial conditions, parametric instabilities are expected
for the low momentum pion modes \cite{4:16}. The energy of the $\sigma$-field
oscillations around $\sigma=f_\pi$ can also be pumped into pionic modes 
through the parametric resonance \cite{4:15,4:17}. In this case, naively only 
modes
with $k\sim \sqrt{\frac{m_\sigma^2}{4}-m_\pi^2}\approx 270~{\mathrm MeV}$ can
be amplified, because oscillation frequency in the $\sigma$-direction is
$m_\sigma=600~{\mathrm MeV}$. However, the energy can be redistributed in the 
long wavelength modes due to nonlinearity.

An interesting example of the parametric instability is Faraday waves 
\cite{4:18} -- surface waves parametrically excited in a vertically vibrating
container of fluid when the vibration amplitude exceeds a certain threshold. 
The resulting standing waves on the fluid surface can form funny intricate 
patterns
\cite{4:19}. Even a more closer analog is given by the induction phenomenon in
quartic Fermi-Pasta-Ulam (FPU) chains \cite{4:20} : the energy, initially 
supplied to a single harmonic mode, remains in this mode over a certain period,
called the induction time, when it is abruptly transferred to other harmonic
modes. The original explanation \cite{4:20} involves nonlinear parametric
instabilities similar to the one described above. It should be mentioned,
however, that our above arguments, in favor of the exponential growth of
fluctuations due to the parametric resonance, are heuristic. A naive 
perturbation theory, implicit in these arguments, is not adequate for such
nonlinear problems. In the case of the FPU chains, it was argued that a more 
correct treatment was provided by shifted-frequency perturbation theory
\cite{4:21} or by Krylov-Bogoliubov-Mitropolsky averaging technique 
\cite{4:22}. We are not aware of similar studies in the context of DCC 
dynamics, but the reality of parametric instabilities is indirectly confirmed
by numerical studies \cite{2:13,4:23}: the observed amplification of 
long-wavelength pionic modes last much longer than expected solely from the 
spinodal instabilities. Besides, the amplification of pionic modes with
$k\approx 270~{\mathrm MeV}$ was clearly demonstrated.

\section{Quantum state of the disoriented chiral condensate}
The eventual decay of the DCC is a quantum process, because one registers
pions and not the classical field. Therefore, the natural question about
the DCC quantum state $|\eta>_{DCC}$ arises. The usual way of quantizing some
classical field configuration is to use coherent states which are eigenstates
of the annihilation operator \cite{5:1}
\begin{equation}
a\,|\alpha>=\alpha \,|\alpha>.
\label{chstate1} \end{equation}
Decomposing
$$|\alpha>=\sum\limits_{n=0}^\infty c_n|n>,\;\;\; |n>=\frac{(a^+)^n}
{\sqrt{n!}}\,|0>$$
and using $a|n>=\sqrt{n}\,|n-1>$, we get from (\ref{chstate1}) the recurrent 
relation $\sqrt{n}\,c_n=\alpha \,c_{n-1}$. Therefore
$$c_n=\frac{\alpha^n c_0}{\sqrt{n!}}$$
and
$$|\alpha>=c_0\,\exp{(\alpha a^+)}\,|0>.$$
However,
$$<\alpha|\alpha>=\sum\limits_n <\alpha|n><n|\alpha>=\sum\limits_n |c_n|^2=
|c_0|^2\exp{(\alpha^*\alpha)},$$
and, therefore, the normalization condition $<\alpha|\alpha>=1$ determines
$c_0$ up to a phase. Finally
\begin{equation}
|\alpha>=\exp{\left(-\frac{\alpha^*\alpha}{2}+\alpha a^+\right)}\,|0>.
\label{chstate} \end{equation}
The generalization of this construction to the DCC classical field 
configuration $f(\vec{x})$ is \cite{5:2} (the isospin indices are suppressed)
\begin{equation}
|\eta>_{DCC}=\exp{\left(-\frac{1}{2}\int d\vec{k} f^*(\vec{k})f(\vec{k})
+\int d\vec{k}f(\vec{k})a^+(\vec{k})\right)}\,|0>,
\label{DCCchs}\end{equation}
where $f(\vec{k})$ is the Fourier transform of the DCC classical field.
 
However, it was argued \cite{5:3,5:4} that the
quantum state of the disoriented chiral condensate is expected to be a 
squeezed state \cite{5:5}, if the parametric amplification mechanism, 
discussed above, indeed plays a crucial role in the DCC formation. To explain 
why, let us consider a one-dimensional,
unit mass, quantum parametric oscillator with the Hamiltonian
$$\hat H(t)=\frac{1}{2}\left [\hat p^2+\omega^2(t)\,\hat x^2 \right ],$$
where the $\hat p$ and $\hat x$ operators are time-independent in the 
Schr\"odinger picture and obey the canonical commutation relation ($\hbar=1$):
$$[\hat x,\hat p]=i.$$
Quantum state vector $|\psi>$ of this oscillator is determined by 
the Schr\"odin\-ger equation
\begin{equation}
i\frac{\partial}{\partial t}\,|\psi>=\hat H(t)\,|\psi>.
\label{Scheq} \end{equation}
Lewis and Riesenfeld gave \cite{5:6} a general method of solving the 
Schr\"odinger equation by using explicitly time-dependent invariants which
are solutions of the quantum Lieuville-Neumann equation
\begin{equation}
\frac{\partial \hat I}{\partial t}+i[\hat H,\hat I]=0.
\label{LNeq} \end{equation}
It turns out that the eigenstates of such a Hermitian invariant 
$\hat I(t)$ are just 
the desired solutions of the Schr\"odinger equation up to some time-dependent
phase factor. Let us demonstrate this remarkable fact \cite{5:6} . 
The eigenvalues of the Hermitian operator $\hat I(t)$ are real. Therefore, 
from (\ref{LNeq}) one easily gets  
\begin{equation}
i<\lambda^\prime|\,\frac{\partial \hat I}{\partial t}\,|\lambda>=
(\lambda -\lambda^\prime)<\lambda^\prime|\hat H(t)|\lambda>,
\label{LN2}\end{equation}
where $|\lambda>$ is an eigenvector of the operator $\hat I(t)$ with the
eigenvalue $\lambda$:
\begin{equation}
\hat I(t)\,|\lambda>=\lambda\,|\lambda>.
\label{LN3} \end{equation}
In particular
\begin{equation}
<\lambda|\,\frac{\partial \hat I}{\partial t}\,|\lambda>=0.
\label{LN4} \end{equation}
By differentiating (\ref{LN3}) with respect to time and taking the scalar
product with $|\lambda^\prime>$, we get
\begin{equation}
<\lambda^\prime|\,\frac{\partial \hat I}{\partial t}\,|\lambda>=
(\lambda -\lambda^\prime)
<\lambda^\prime|\,\frac{\partial }{\partial t}\,|\lambda>+\;
\delta_{\lambda\lambda^\prime}\,\frac{\partial \lambda}{\partial t}.
\label{LN5} \end{equation}
For $\lambda=\lambda^\prime$, we get from (\ref{LN5}) and (\ref{LN4})
$$\frac{\partial \lambda}{\partial t}=0.$$
That is the eigenvalues of the operator $\hat I(t)$ are time-independent
(as it should be for the invariant operator). From (\ref{LN5}) and 
(\ref{LN2}) one gets
$$(\lambda -\lambda^\prime)
<\lambda^\prime|\,i\frac{\partial }{\partial t}\,|\lambda>=
(\lambda -\lambda^\prime)<\lambda^\prime|\hat H(t)|\lambda>$$
and, therefore
\begin{equation}
<\lambda^\prime|\,i\frac{\partial }{\partial t}\,|\lambda>=
<\lambda^\prime|\hat H(t)|\lambda>,\;\;\;{\mathrm if}\;\; \lambda^\prime\ne
\lambda.
\label{LN6}\end{equation}
If (\ref{LN6}) is also satisfied for the diagonal matrix elements ($\lambda=
\lambda^\prime$), then we can immediately deduce that $|\lambda>$ is a 
solution of the Schr\"odinger equation. But this may not be the case for our
particular choice of eigenvectors. Nevertheless, in this case we can still
adjust the phases of the eigenvectors in such a way that the new eigenstates
$$|\lambda>^\prime=e^{i\theta_\lambda(t)}\,|\lambda>,\;\;
\theta_\lambda(0)=0,$$
insure the validity of (\ref{LN6}) for all $\lambda,\,\lambda^\prime$. All
what is needed is to choose the time-dependent phases $\theta_\lambda(t)$ 
in such a way that one has
$$<\lambda|\,e^{-i\theta_\lambda}\left (i\frac{\partial}{\partial t}\right )
e^{i\theta_\lambda}\,|\lambda>= <\lambda |\hat H(t)|\lambda>,$$
or
$$\frac{d\theta_\lambda}{dt}=<\lambda |\,i\frac{\partial}{\partial t}-
\hat H(t)\,|\lambda>.$$
Therefore
$$\theta_\lambda(t)=\int\limits_0^t <\lambda |\,i\frac{\partial}{\partial 
\tau}-\hat H(\tau)\,|\lambda>\,d\tau.$$
To summarize, we can take any set $|\lambda>$ of the eigenstates of the 
invariant operator $\hat I(t)$ and express the general solution of the
Schr\"odinger equation (\ref{Scheq}) as a linear combination
\begin{equation}
|\psi>=\sum_\lambda C_\lambda \exp{\left \{ i 
\int\limits_0^t <\lambda |\,i\frac{\partial}{\partial
\tau}-\hat H(\tau)\,|\lambda>\,d\tau\right\}}\,|\lambda>,
\label{gensol} \end{equation}
with some time-independent coefficients $C_\lambda$.

To find the invariant $\hat I(t)$, let us firstly construct its classical 
counterpart $I(t)$ by the simple and transparent method of Eliezer and Gray
\cite{5:7}. The equation of motion
$$\ddot x+\omega^2(t)\,x=0$$
can be viewed as the $x$-projection of a two-dimensional auxiliary motion
governed by the equation
\begin{equation}
\ddot {\vec{r}}+\omega^2(t)\,\vec{r}=0,
\label{auxmot} \end{equation}
where $\vec{r}=x\,\vec{i}+y\,\vec{j}$. In the polar coordinates
$$x=\rho\cos{\varphi},\;\;\;y=\rho\sin{\varphi},$$
and
$$\ddot {\vec{r}}=\left (\ddot \rho-\rho\,{\dot \varphi}^2\right )
{\vec{e}}_\rho+
\left (\rho\,\ddot\varphi+2\dot\rho\,\dot\varphi\right ){\vec{e}}_\varphi,$$
where the unit basic vectors are
$${\vec{e}}_\rho=\cos{\varphi}\,\vec{i}+\sin{\varphi}\,\vec{j},\;\;
{\vec{e}}_\varphi=-\sin{\varphi}\,\vec{i}+\cos{\varphi}\,\vec{j}.$$
Therefore, (\ref{auxmot}) in the polar coordinates takes the form
$$\ddot\rho-\rho\,{\dot \varphi}^2+\omega^2(t)\rho=0,\;\;\;
\rho\,\ddot\varphi+2\dot\rho\,\dot\varphi=\frac{1}{\rho}\,
\frac{d(\rho^2\dot\varphi)}{dt}=0.$$
The second equation implies
$$\rho^2\dot\varphi=L=const.$$
This is nothing but the conservation of the angular momentum for the 
auxiliary planar motion, when the first equation can be rewritten as
$$\ddot\rho+\omega^2(t)\rho=\frac{L^2}{\rho^3}.$$
Let us now remark that (for unit mass)
$$p=\dot x=\dot\rho\cos{\varphi}-\rho\dot\varphi\sin{\varphi}=
\frac{\dot\rho x-L\sin{\varphi}}{\rho}.$$
Therefore,
$$L\sin{\varphi}=\dot\rho x-\rho p,\;\;\; L\cos{\varphi}=\frac{L}{\rho}x,$$
and
$$\frac{L^2}{\rho^2}x^2+(\rho p-\dot\rho x)^2=L^2=const.$$
The Ermakov-Lewis invariant \cite{5:8,5:9} corresponds to the particular case
when the angular momentum $L$ has a unit value
$$I(t)=\frac{1}{2}\left [ \frac{x^2}{\rho^2}+(\rho p-\dot\rho x)^2\right ].$$
It is straightforward to check that its quantum counterpart
\begin{equation}
\hat I(t)=\frac{1}{2}\left [ \frac{\hat x^2}{\rho^2}+(\rho \hat p-
\dot\rho \hat x)^2\right ]
\label{ELinv} \end{equation}
obeys the quantum Lieuville-Neumann equation (\ref{LNeq}) if the auxiliary 
function $\rho(t)$ satisfies the Ermakov-Milne-Pinney equation \cite{5:10} 
\begin{equation}
\ddot\rho+\omega^2(t)\rho=\frac{1}{\rho^3}\;.
\label{rhoeq} \end{equation}
To find eigenvectors of the operator $\hat I(t)$, let us note that
\begin{equation}
\hat I(t)=b^+(t)b(t)+\frac{1}{2}\;,
\label{Icanform} \end{equation}
where we have introduced time-dependent ``creation'' and ``annihilation'' 
operators
\begin{equation}
b(t)=\frac{1}{\sqrt{2}}\left [ \frac{\hat x}{\rho}+i(\rho\,\hat p-\dot \rho\,
\hat x)\right ],\;\;\; b^+(t)=\frac{1}{\sqrt{2}}\left [ \frac{\hat x}{\rho}
-i(\rho\,\hat p-\dot \rho\,\hat x)\right ].
\label{bboper} \end{equation}
It can be immediately checked that one indeed has the canonical commutation 
relation
\begin{equation}
[b(t),b^+(t)]=1.
\label{bbcomm} \end{equation}
Equations (\ref{Icanform}) and (\ref{bbcomm}) indicate that the eigenvectors
of $\hat I(t)$ are $b$-number states
\begin{equation}
|n;b>=\frac{(b^+(t))^n}{\sqrt{n!}}\,|0;b>,
\label{bnstates} \end{equation}
where the $b$-vacuum state is defined by the condition
\begin{equation}
b(t)\,|0;b>=0.
\label{bvacuum} \end{equation}
Using these eigenvectors, one can construct the solution of the
Schr\"odinger equation, as described above. In the DCC case, however, it is
preferable to express the quantum state vector in terms of the pion creation
and annihilation operators $a^+,\,a$. So we need a relation between two sets
of the creation-annihilation operators $a^+,\,a$ and $b^+,\,b$ (the Bogoliubov
transformation). The pionic modes correspond to the late time asymptotes 
$\omega(t)\to\omega(\infty)=\omega_0$. The solution of the 
Ermakov-Milne-Pinney equation for $\omega(t)=\omega_0=const$ is
$$\rho=\frac{1}{\sqrt{\omega_0}}\;.$$
Therefore, the ``pionic'' creation-annihilation operators have the standard 
form
\begin{equation}
a=\frac{1}{\sqrt{2\omega_0}}\left [ \omega_0\,\hat x+i\hat p\right ],\;\;\; 
a^+=\frac{1}{\sqrt{2\omega_0}}\left [ \omega_0\,\hat x-i\hat p\right ].
\label{aaoper} \end{equation}
The comparison of (\ref{aaoper}) and (\ref{bboper}) gives the desired 
Bogoliubov transformation
\begin{equation}
b(t)=\alpha(t)\,a+\beta^*(t)\,a^+,\;\;\; b^+(t)=\beta(t)\,a+\alpha^*(t)\,a^+,
\label{Btrans} \end{equation}
where
\begin{eqnarray} & &
\alpha(t)=\frac{1}{2\sqrt{\omega_0}}\left [ \frac{1}{\rho(t)}+\omega_0\,
\rho(t)-i\,\dot\rho(t)\right], \nonumber \\ & &
\beta(t)=\frac{1}{2\sqrt{\omega_0}}\left [ \frac{1}{\rho(t)}-\omega_0\,
\rho(t)+i\,\dot\rho(t)\right].
\label{Bcoeff} \end{eqnarray}  
It can be checked that
$$|\alpha(t)|^2-|\beta(t)|^2=1.$$
Therefore, up to an irrelevant common phase, we can take
\begin{equation}
\alpha(t)=\cosh{r(t)},\;\;\; \beta(t)=e^{i\delta(t)}\sinh{r(t)}.
\label{albtpar} \end{equation}
The Bogoliubov transformation (\ref{Btrans}) can be viewed as an unitary
transformation
\begin{equation}
b=\hat S(z)\,a\,\hat S^+(z),\;\;\; b^+=\hat S(z)\,a^+\,\hat S^+(z),
\label{Utrans} \end{equation}
where $\hat S(z)$ is the so-called squeezing operator \cite{5:5}
\begin{equation}
\hat S(z)=\exp{\left [\frac{1}{2}\left (z\,a^+a^+-z^*\,aa\right )\right ]},
\;\; z=r\,e^{i(\delta+\pi)}.
\label{sqzoper} \end{equation}
Indeed, using the Campbell-Hausdorf formula
$$e^{\hat B}\hat Ae^{-\hat B}=\hat A+[\hat B,\hat A]+
\frac{1}{2!}[\hat B,[\hat B,\hat A]]+\frac{1}{3!}[\hat B,[\hat B,[\hat B,
\hat A]]]+\ldots $$
and summing up the resulting infinite series, one can check that 
(\ref{Utrans}) and (\ref{Btrans}) are equivalent, if $\alpha$ and $\beta$ 
are given by (\ref{albtpar}).

Therefore, the DCC quantum state is expected to have the form
$$|\eta>_{DCC}=\hat S(z) |\Psi_0>,$$
where $|\Psi_0>$ is the DCC initial state before the onset of the parametric
amplification. If $|\Psi_0>$ is a coherent state, then the resulting
$|\eta>_{DCC}$ state will be the one called the squeezed state 
\cite{5:5}. The actual parameters of this state (for example $z$) are 
determined by the initial conditions and are hard (if not impossible) to 
estimate from the theory alone.

Let us now recall the isospin, for a moment. The isospin generators are
$$\vec{I}=\int\vec{\pi}(x)\times\dot{\vec{\pi}}(x)\,d\vec{x}.$$
The classical pion field $\vec{\pi}$ of the DCC points in some fixed direction,
$\vec{n}$ in the isospace. If its time derivative $\dot{\vec{\pi}}$ also 
points in the same direction then $\vec{I}=0$ and we will have an isosinglet
state. One can expect such a situation in $\rho-\rho$ collisions (which 
dominates in the $\gamma\gamma\to hadrons$ cross section), because the
``vacuum cleaning'' effect, which precedes the DCC formation, in this case is
mainly due to two colliding isospin blind gluon walls. It is easy to construct
an isosinglet squeezed state by just exponentiating the apparently isoscalar 
operator
$$-\sum\limits_{i=1}^3 a_i^+a_i^+=2a_+^+a_-^+-a_3^+a_3^+\, ,$$
where $a_{\pm}^+=\frac{\mp 1}{\sqrt{2}}(a_1^+\pm ia_2^+)$ are the charged-pion 
creation operators. The resulting squeezed state is \cite{5:4}
$$|\Psi>=N\,\exp{\left \{\frac{\alpha}{2}\left (2a_+^+a_-^+-a_3^+a_3^+\right )
\right\}}\,|0>.$$
Well, this expression does not seem, at first glance, to correspond to the 
canonical form (\ref{sqzoper}) of the squeezing operator. In fact, it 
indeed gives a squeezed state. This is clear from the following normal-ordered
form of the squeezing operator \cite{5:5}
\begin{equation}
S(re^{i\theta})=N\,\exp{\left(\frac{\alpha}{2}\,a^+a^+\right)}\sum
\limits_{n=0}^\infty\frac{({\mathrm sech}\,r-1)^n}{n!}(a^+)^n(a)^n\, 
\exp{\left(-\frac{\alpha}{2}\,aa\right)},
\label{nrmsqz} \end{equation}
where $\alpha=e^{i\theta}\tanh{r}$ and
$$N=\frac{1}{\sqrt{\cosh{r}}}=\left (1-|\alpha|^2\right)^{1/4}.$$

What is the probability $P(m;n)$ that $|\Psi>$ decays by producing in total 
$2n$ pions and among them $2m$ neutral pions and equal
numbers of positively and negatively charged pions? According to the standard
rules of quantum mechanics
$$P(m;n)=|<m;n|\Psi>|^2,$$
where the normalized state $|m;n>$ is defined through
$$|m;n>=\frac{1}{\sqrt{(2m)!}}\,\frac{1}{(n-m)!}\,(a_+^+a_-^+)^{(n-m)}
\,(a_3^+)^{2m}\,|0>.$$
However,
$$|\Psi>=N\,\sum\limits_{k=0}^\infty \frac{(\alpha/2)^k}{k!}\,
\left (2a_+^+a_-^+-a_3^+a_3^+\right )^k\,|0>=$$ $$=N\,\sum\limits_{k=0}^\infty 
\frac{(\alpha/2)^k}{k!}\,\sum\limits_{l=0}^k C_k^l\,2^{k-l}\,\sqrt{(2l)!}\,
(l-k)!\,|l;k>.$$
Therefore,
$$P(m;n)=\left |N\,\frac{(\alpha/2)^n}{n!}\,C_n^m\,2^{n-m}\,\sqrt{(2m)!}\,
(n-m)!\right |^2=N^2\,|\alpha|^{2n}\,\frac{(2m)!}{(m!\,2^m)^2}\,.$$
One can prove by induction in $n$ that
$$\sum\limits_{m=0}^n \frac{(2m)!}{(m!\,2^m)^2}=\frac{(2n+1)!}{(n!\,2^n)^2}.$$
This enables one to express $P(m;n)$ as a product of two probabilities:
$$P(m;n)=P_1(n)P_2(m;n),$$
where $P_1(n)$ is the probability that one will find the total number of $2n$ 
pions after the state $|\Psi>$ decays
$$P_1(n)=N^2\,|\alpha|^{2n}\,\frac{(2n+1)!}{(n!\,2^n)^2}.$$
Note that
$$\sum\limits_{n=0}^\infty \frac{(2n+1)!}{(n!\,2^n)^2}\,|\alpha|^{2n}=
1+\frac{3}{2}\,|\alpha|^2+\frac{3\cdot 5}{2\cdot 4}\,|\alpha|^4+
\frac{3\cdot 5\cdot 7}{2\cdot 4\cdot 6}\,|\alpha|^6+\cdots =(1-|\alpha|^2)^
{-3/2}.$$
Therefore, the normalization coefficient $N$ should be
$N=(1-|\alpha|^2)^{3/4}$, and this is exactly what is expected from 
(\ref{nrmsqz}) for the product of three properly normalized single-mode 
squeezed states of Cartesian pions.

More interesting for us is the second factor, $P_2(m;n)$, the probability that
one finds $2m$ neutral pions in such a $2n$-pion final state \cite{4:9,5:HS}:
$$P_2(m;n)=\frac{(n!\,2^n)^2}{(2n+1)!}\,\frac{(2m)!}{(m!\,2^m)^2}.$$
In fact, $P_2(m;n)$ is a particular case of the Poly\'{a} distribution 
\cite{5:11}. If $m$ and $n$ are both large, one can use the Stirling
formula
$$n!\approx \left(\frac{n}{e}\right)^n\,\sqrt{2\pi n}$$
to get
$$P(m;n)\approx \frac{1}{2n}\,\frac{1}{\sqrt{m/n}}.$$
Therefore the same inverse-square-root distribution (\ref{ISQRTD}) is 
recovered in the continuum limit.

A few more words about coherent and squeezed states, in order to make them 
more familiar. If $[\hat A,\hat B]$ is a $c$-number, then \cite{5:12} 
$$e^{\hat A+\hat B}=e^{\hat A}\,e^{\hat B}\,e^{-\frac{1}{2}\,
[\hat A,\hat B]}.$$
Using this theorem, we get
$$\exp{[\alpha\,a^+-\alpha^*\,a]}=\exp{\left [ -\frac{1}{2}\alpha^*\,\alpha
\right ]}\,\exp{(\alpha\, a^+)}\,\exp{(-\alpha^*\, a)}.$$
But $\exp{(-\alpha^*\, a)}\,|0>=|0>$. Therefore, the coherent state $|\alpha>$
can be generated by the unitary displacement operator: $|\alpha>=\hat D(
\alpha)\,|0>$, where
\begin{equation}
\hat D(\alpha)=\exp{[\alpha\, a^+-\alpha^*\, a]}.
\label{disoper} \end{equation}
Now let us consider the ground state for a harmonic oscillator (we have 
abandoned the unit mass restriction but will still keep $\hbar =1$):
$$\Psi_0(x)\equiv <x|0>=\left [2\pi\sigma_0^2\right ]^{-1/4}\,
\exp{\left [-\left (\frac{x}{2\sigma_0}\right )^2\right ]},\;\;\;
\sigma_0^2=\frac{1}{2m\omega},$$
and calculate the effect of the displacement operator on it. For the harmonic
oscillator 
$$a=\frac{1}{\sqrt{2m\omega}}\,[m\omega\,\hat x+i\,\hat p],\;\;\;
a^+=\frac{1}{\sqrt{2m\omega}}\,[m\omega\,\hat x-i\,\hat p].$$
Hence,
$$\alpha\,a^+-\alpha^*\,a=ip_0\,\hat x-ix_0\,\hat p,$$
where
$$x_0=\sqrt{\frac{2}{m\omega}}\,{\mathrm Re}\,\alpha,\;\;\;
p_0=\sqrt{2m\omega}\,{\mathrm Im}\,\alpha.$$
In the coordinate representation $\hat p=-i\frac{\partial}{\partial x}$;
therefore ,
$$\Psi_{cs}(x)\equiv <x|\alpha>=\exp{\left [ip_0x-x_0\frac{\partial}
{\partial x}\right ]}\, \Psi_0(x)=$$
$$=\left [2\pi\sigma_0^2\right ]^{-1/4}\,
\exp{\left [-\left (\frac{x-x_0}{2\sigma_0}\right )^2+ip_0x 
-i\,\frac{x_0\,p_0}{2} \right ]}.$$
As we see, for harmonic oscillator, the coherent state is a Gaussian which
is displaced from the origin by $x_0$. It has the ground state width $\sigma_0$
and a phase linearly dependent on the 
position $x$. Schr\"{o}dinger discovered
\cite {5:13} such a state as early as 1926 while seeking ``unspreading wave 
packets''.

What about the squeezed vacuum state $\Psi_{s0}=\hat S(z)\,\Psi_0$? Note that
(for simplicity we will take $z=r$ to be real)
$$a^+a^+-aa=-i(\hat x\,\hat p+\hat p\,\hat x\,).$$
Then
$$\frac{-ir}{2}\,[\hat x\,\hat p+\hat p\,\hat x,\hat x]=-r\hat x,\;\;\;
\frac{-ir}{2}\,[\hat x\,\hat p+\hat p\,\hat x,\hat p]=r\hat p,$$
and the Campbell-Hausdorf formula will give
$$\hat S(r)\,\hat x\,\hat S^+(r)=e^{-r}\,\hat x,\;\;\;
\hat S(r)\,\hat p\,\hat S^+(r)=e^{r}\,\hat p.$$
Therefore
$$\hat S(r)\,\hat H\,\hat S^+(r)=\frac{e^{2r}\hat p^2}{2m}+\frac{m\omega^2
e^{-2r}\hat x^2}{2}=-\frac{1}{2m}\,\frac{\partial^2}{\partial(e^{-r}x)^2}+
\frac{m\omega^2}{2}\,(e^{-r}x)^2,$$
and comparing the equations which $\Psi_{s0}$ and $\Psi_0$ are satisfying
$$\hat S(r)\,\hat H\,\hat S^+(r)\,\Psi_{s0}=\frac{\omega}{2}\,\Psi_{s0},\;\;\;
\hat H \,\Psi_0=\frac{\omega}{2}\,\Psi_0,$$
we conclude that
$$\Psi_{s0}(x)=\Psi_0(e^{-r}x)=\left [2\pi\sigma_0^2\right ]^{-1/4}\,
\exp{\left [-\left (\frac{x}{2e^r\sigma_0}\right )^2\right ]}.$$
In a more general case, the squeezed state $\Psi_{ss}=\hat D(\alpha)\hat S(r)\,
\Psi_0$ is a Gaussian
$$\Psi_{ss}(x)=\left [2\pi\sigma_0^2\right ]^{-1/4}\,
\exp{\left [-\left (\frac{x-x_0}{2\sigma}\right )^2+ip_0x 
-i\,\frac{x_0\,p_0}{2} \right ]},$$
where
$$\sigma=e^r\sigma_0.$$
Therefore, the squeezed state differs from the coherent state only by squeezing
in (if $e^r<1$), or squeezing out (if $e^r>1$) the width of the ground state
Gaussian. The momentum-space wave function is squeezed oppositely.

For further discussion see \cite{5:5,5:14}. Let us note that squeezed states
were also discovered long ago (in 1927) by Kennard \cite{5:12}. Both the
Schr\"{o}dinger and Kennard works had little impact, and both coherent
and squeezed states, which are now cornerstones of quantum optics, were
rediscovered after decades. As Nieto remarks,   
``To be popular in physics you have to either be good or lucky.
Sometimes it is better to be lucky. But if you are going to be good,
perhaps you should not  be too good''.

As the last, but not the least, question of this section, let us ask whether
a quantum state of the disordered chiral condensate $|\eta>_{DCC}$ may be
produced without any intermediate phase transitions altogether,
through the quantum reaction
$\gamma\,\gamma\to |\eta>_{DCC}$. The functional integral methods and
stationary phase approximation (semiclassical approximation) are natural
tools to study the scattering amplitudes between initial wave packet states
and certain final coherent states \cite{5:15, 5:16}. In this article we 
are not really interested in actual calculations of this type. Our aim at the
beginning is more humble -- to provide some arguments that such a quantum 
transition is indeed possible and interesting. So we will consider an 
oversimplified toy model with the (second quantized) Hamiltonian
$$\hat H=\omega\,a^+a+2\omega\,b^+b+g\,(ba^+a^++b^+aa),$$
with $2\omega=m_\pi$. Here the $b$-mode mimics neutral pions, $a$-mode -- 
photons,
and the interaction term with $g=\frac{\alpha\,m_\pi^2}{\pi\,f_\pi}$ imitates
the $\pi^0\gamma\gamma$ interaction due to axial anomaly. Such Hamiltonians 
are used in quantum optics to study  the second-harmonic generation 
\cite{5:17, 5:18}. We further assume that all available initial energy, 
$\sqrt{s}$, is accumulated in the $a$-mode. That is, the assumed initial 
$a$-mode occupancy is
$$N_a=\frac{\sqrt{s}}{\omega}=\frac{2\sqrt{s}}{m_\pi}\, .$$
It can be easily checked that $[\hat H_0,\hat H_{int}]=0$, where
$$\hat H_0=\omega\,a^+a+2\omega\,b^+b,\;\;\;
\hat H_{int}=g\,(ba^+a^++b^+aa).$$
This means that both $\hat H_0$ and $\hat H_{int}$ are constants of motion.
We can simply forget about $\hat H_0$, because it just gives an 
irrelevant common phase factor $\exp{(iN_a\omega t)}$ in the evolution 
operator. Hence the nontrivial part of the initial-state evolution is given by
the relation
$$|\Psi(t)>=e^{i\hat H_{int}t}\,|N_a;0>.$$
However, $N_a\gg 1$; therefore, as far as the $b$-mode initial development is 
concerned, we can replace the $a$ and $a^+$ operators in $\hat H_{int}$ by 
the $c$-numbers
$\alpha$ and $\alpha^*$, at that $|\alpha|^2=N_a$. In this approximation
\begin{equation}
|\Psi(t)>=e^{\beta\,b^+-\beta^*\, b}\,|0>,\;\;\; \beta=ig\,\alpha^2\, t.
\label{Psib} \end{equation}
As we see, a coherent state of $b$-quanta is being formed. The mean number of
quanta in this state grows with time (until the approximation considered 
breaks down) as follows:
\begin{equation}
N_b=|\beta|^2=(gN_at)^2.
\label{Nbt} \end{equation}
To estimate the terminal time, let us note that the variance of the number of
quanta in the coherent state (\ref{Psib}) also equals $|\beta|^2$ (see, for
example, \cite{5:14}). Therefore, the development time for the coherent state
(\ref{Psib}) can be estimated from the energy-time uncertainty relation
$$\sqrt{N_b}\,m_\pi t\sim 1.$$
Using this estimation, we finally get from (\ref{Nbt}) the relation
$$\sqrt{s}=\frac{\pi\,N_b}{2\,\alpha}\,f_\pi.$$
Therefore, for example, $N_b \sim 100$ implies $\sqrt{s}\sim 2~{\mathrm 
{TeV}}$.

Having in mind a very crude and heuristic nature of our arguments, we admit 
that we may easily be wrong by an order of magnitude. Nevertheless, the main 
indications of the above exercise that the axial anomaly can lead to the 
generation of a pionic coherent state in gamma-gamma collisions, and that the 
efficient generation requires not fantastically high center-of-mass energies
certainly seems interesting and deserves further study.

\section{Concluding remarks}
"Have no fear of perfection -- you'll never reach it" -- Salvador Dali once 
remarked. At the end of our enterprise we reluctantly realized how true 
the second half of this quotation is. Therefore, we abandon an unrealistic 
dream to produce the perfect review of the disoriented chiral condensate 
and try to finish here. It is important to finish at right time, is not
it? One American general began his talk with the following
sentence: ``My duty is to speak, and your duty is to listen. And I hope to
end my duty before you end yours''. We hope that the reader has not yet
end his duty, because there is one topic which should be touched a little 
before we finish.

The disoriented chiral condensate is a very attractive idea, and it has some
solid theoretical support behind it, as we tried to demonstrate above. But 
are there any experimental indications in favor of its real existence?
In fact there are some exotic cosmic ray events, called Centauros, where one
may suspect the DCC formation. Centauros were discovered in high-mountain
emulsion chamber cosmic ray experiments \cite{6:1}. Typically, the detectors
used in such experiments consist of the upper and lower chambers separated by
the carbon target. Each chamber is a sandwich of the lead absorber and the 
sensitive layers. The normal cosmic ray event is usually generated by the
primary interaction at about $500-1000~{\mathrm m}$ altitude above the 
detector apparatus. About one third of the products of the primary 
interaction are neutral pions. Each neutral pion decays into two $\gamma$ 
quanta. Therefore, roughly one $\gamma$ quantum is expected per a charged 
particle in the primary interaction. When the interaction products reach the 
upper chamber the numbers of electronic and photonic secondaries are much 
increased through the electromagnetic shower formation. Therefore, the upper
chamber usually detects several times more particles than the lower chamber,
because the electromagnetic component is strongly suppressed by the carbon 
layer, leaving mainly the hadronic component to be detected by the lower
chamber. A big surprise was the discovery of events with the contrary 
situation.
Such events were named ``Centauros'' because it was not possible to guess
their lower parts from the upper ones. 

The first Centauro was observed in 1972 at the Chacaltaya high mountain 
laboratory
\cite{6:1}. It was initiated by the primary interaction at a relatively low 
altitude, at only $(50\pm 15)~{\mathrm m}$ above the detector. Therefore, the
event was very clean, that is was almost not distorted by electromagnetic and
nuclear cascades in the atmospheric layer above the chambers. After correcting
for the hadron detection efficiency and for the influence of the secondary 
atmospheric interactions, the event can be interpreted as the production of
only one electromagnetic ($e/\gamma$) particle and 74 hadrons with the total
interaction energy $\approx 330~\mathrm{TeV}$.

Afterwards some more Centauros were found. Namely \cite{6:1}, the Chacaltaya
experiment observed 8 unequivocal Centauros, and two experiments at Pamir
found 3 and 2 more Centauros. But no clean Centauros were found in Kanbala
and Fuji experiments -- the puzzle which still remains a mystery \cite{6:1}.
However, if the definition of Centauro is somewhat relaxed and all hadron-rich
species are considered, then such Centauro-like anomalies constitute about
20\% of events with the total visible energy $\ge 100~\mathrm{TeV}$ \cite{6:1},
hence they are by no means rare phenomena at such high energies.

Of course the DCC formation is a candidate explanation of Centauro events. 
However such explanation is not without difficulties \cite{6:1}. For example, 
the 
large transverse momenta observed in the Centauro events is difficult to 
explain in the DCC scenario. It is not also evident that the Centauro hadrons
are pions. If they are mostly baryons instead, then an alternative explanation
may be provided by strangelets \cite{6:2}. 

Let us note, however, that if one takes the DCC explanation of Centauros
seriously, some predictions immediately follow. First of all, it may happen
that the DCC domain is produced in the cosmic-ray interactions with 
significant transverse velocity. In this case the coherent pions from the DCC
decay will constitute ``coreless jet'' in the laboratory frame, with pions in
the jet having small ($<100~\mathrm{MeV}$) relative transverse momenta 
\cite{3:6}. Interestingly, such hadron-rich events, called Chirons, were 
really observed in both Chacaltaya and Pamir experiments \cite{6:1}.

If the DCC is aligned along the $\pi^0$-direction in isospace, then a 
particular
anti-Centauro event is expected with neutral pion fraction $f$ close to 
unity. For example, one has not very small probability that the neutral pion 
fraction is in the interval $0.99\le f \le 1$:
$$P(0.99\le f \le 1)=\int\limits_{0.99}^{1.0}\frac{df}{2\sqrt{f}}\approx
0.5\% .$$ 
No such anti-Centauro events were observed in the Chacaltaya and Pamir 
experiments. However, some anti-Centauros were reported in the 
Japanese-American
JACEE experiment, in which the emulsion chambers were flown near the top of 
the atmosphere by balloons \cite{6:1}. By the way, this experiment has not 
seen any Centauro events -- another mystery puzzle of this cosmic ray Centauro
business.

Of course Centauro-like events were searched in accelerator experiments 
\cite{6:1}. The first searches have been performed by UA1 and UA5 experiments
at CERN even before the DCC idea was suggested. Both experiments found no
Centauro candidates in the central rapidity region. 

The estimated average energy of Cosmic-ray Centauro events is about $1740~
\mathrm{TeV}$ \cite{6:1}. If  Centauros are formed in nucleon-nucleon 
collisions, this energy threshold translates into $\sqrt{s}\approx 1.8~
\mathrm{TeV}$ in the c.m. frame -- roughly the Tevatron energy. This 
observation maybe explains the failure of UA1 and UA5 experiments, where the
maximal available energy was $\sqrt{s}\approx 0.9~\mathrm{TeV}$, and makes
Fermilab experiments more attractive in this respect. However, one has to
bear in mind that there is a crucial kinematic difference between 
cosmic-ray and collider experiments \cite{6:1}: the cosmic-ray experiments
generally detect particles from the fragmentation rapidity region whereas the
collider experiments are mainly focused on the central rapidity region. 
Therefore, the fact of observation of cosmic-ray Centauros does not 
automatically guarantee that these beasts can be found in Tevatron 
experiments.

A small test experiment Mini-Max (T-864) \cite{6:3} at Tevatron was specially
designed for a search of DCC in the forward region. The results of this
experiment \cite{6:4} are consistent with the generic production mechanism and
show no evidence of the presence of DCC. Despite this failure to find DCC,
the Mini-Max experiment was an important benchmark. It was demonstrated for 
the first time that it was  possible to work in the very forward region with
severe background conditions. Much was learned in both detector operation 
and data analysis which should prove useful in future more elaborate
efforts of this kind.

Central rapidity region Centauros were searched in the CDF experiment at 
Tevatron
with negative result \cite{6:1}. Another major Tevatron detector D0 is also
suitable for such searches, as the Monte Carlo study shows \cite{6:1}.

A serious effort to study possible DCC formation in heavy ion collisions was
undertaken in the CERN SPS fixed target experiment WA98 \cite{6:5}. Again no
DCC signal was found in the central $158\cdot A$ GeV Pb+Pb collisions.
 
A majority of future heavy ion experiments at RHIC and LHC have plans to look
for the Centauro phenomena \cite{6:1}. The kinematic conditions at which 
cosmic ray
Centauros are produced will be accessible at RHIC. Therefore the corresponding
experiments (PHOBOS, STAR, PHENIX and BRAHMS) are very interesting in light of
Centauro investigation. At LHC the energy accessible in Pb+Pb central 
collisions will be much higher than the expected threshold energy for the 
Centauro production. Besides, Pb+Pb collisions at LHC will produce a very 
high baryon number density in the forward rapidity region. To study the novel
phenomena expected in such high baryochemical potential environment, the 
CASTOR detector, as the part of the ALICE experiment, was designed \cite{6:6}.
Its main goal is the Centauro and strangelet search in the very forward 
rapidity region in nucleus-nucleus collisions.

We believe that future photon-photon colliders are also good places to look
for the DCC production. Some hints were given above that the DCC formation
conditions might be even more favorable at photon-photon colliders rather 
than at
proton-proton (or proton-antiproton) colliders. Here we present one more 
argument of this kind which deals with the very different roles played 
by gluons in mesons and baryons. Mesons can be considered as a quark-antiquark
pair connected by a gluon string (flux tube). Therefore, the gluon field 
configuration in mesons is, in some sense, topologically trivial. In baryons 
one has a quite different picture \cite{6:7,6:8,6:9}. According to the common
wisdom, baryons are three-quark bound states. In high energy $pp$ or heavy
ion collisions the valence quark distributions in the projectiles will be
significantly Lorentz-contracted because a typical fraction of the proton's
momentum carried by a valence quark is $\sim 1/3$. Therefore, one expects that
the constituent quarks of the colliding protons will not have enough time to
interact significantly during the high-energy collisions and hence it is 
difficult to stop them. If the baryon number of the projectile is associated
with their valence quarks, which is the naive expectation, then the ready 
prediction from the above given collision picture will be that the baryon 
number flow should be concentrated at large positive and negative rapidities,
with a nearly zero net baryon number at central rapidities. Surprisingly, this
is not the case supported by experiments. On the contrary, experiments suggest
that the valence quarks do not carry the proton's baryon number and the flow
of the baryon number can be separated from the flow of the valence quarks 
\cite{6:8,6:9}! But then what is the mysterious fourth constituent of the 
proton which traces its baryon number? In QCD the baryon is represented by
a gauge invariant, non-local, color singlet operator. In fact, the gauge 
invariance constraint severely restricts the possible forms of such composite
operator, leaving only one possibility ($\alpha,\beta,\gamma$ are the color 
indices, the flavor indices are suppressed for simplicity) :
$$B=\epsilon_{\alpha\beta\gamma}\left [\hat T (x_1,x) q(x_1) \right ]^\alpha
\left [\hat T (x_2,x) q(x_2) \right ]^\beta 
\left [\hat T (x_3,x) q(x_3) \right ]^\gamma.$$
Here $\hat T(x_i,x)$ is the open string operator (the Wilson line), or 
parallel transporter of the quark field $q(x_i)$ from the point $x_i$ to the 
point $x$, where the three strings join. This string operator is an analog
of the well known Aharonov-Bohm phase in QED and is given by the path-ordered
exponent
$$\hat T (x_i,x)=P\,\exp{\left (ig\int\limits_{x_i}^x A_\mu\,dx^\mu
\right )},\;\;\; A_\mu=A_\mu^a\frac{\lambda_a}{2}.$$
Therefore, the gluon strings (flux tubes) inside a baryon have nontrivial 
Y-shaped topology and one finds a novel object there -- the string junction.
This string junction is just the fourth constituent of the baryon which traces
its baryon number \cite{6:8}. The string junction can be more easily stopped 
in the high-energy collisions, because, being formed from the soft gluons, 
it is not Lorentz-contracted and always has enough time to interact.

Now we have the following picture of the high-energy $pp$ collisions 
\cite{6:8}: the valence quarks are stripped-off and produce jets in the
fragmentation regions. In some events, one or both of the string junctions 
are stopped in the central rapidity region producing a violent gluon sea
containing one or two twists. On the contrary, no twists are expected in
the gluon sea produced by high-energy photon-photon ($\rho\rho$) collisions.
We believe that the latter situation is more favorable for the Baked Alaska
scenario and, therefore, for the DCC production through this mechanism.

As a final remark, let us note that the DCC formation is just one interesting
collective effect expected in high-energy collisions. Other exotic phenomena
are also worth to be searched. Let us mention a few: the possible formation of 
the pion and eta strings during the chiral phase transition \cite{6:10}, 
creation of the parity and $CP$ violating metastable vacuum bubbles 
\cite{6:11}, production of QCD Buckyballs -- femtometer scale gluon junction
networks (QCD analog of the nanoscale carbonic Fullerenes) \cite{6:9}.
Vacuum engineering at photon colliders promises to be an exciting adventure
and we suspect that one may encounter ``totally unexpected'' new phenomena
during such exploration:
``There are more things in Heaven and Earth, Horatio, than are dreamt of in
your philosophy'' \cite{6:WS}.

\section*{Acknowledgments}
We are grateful to Valery Telnov for discussions. Support from  INTAS 
grants 00-00679 and  00-00366 is acknowledged.
We are grateful to G.G. Sandukovskaja for help.

\section*{Appendix: the Baked Alaska recipe}
Here we reproduce the recipe from \cite{A}.

\subsection*{Ingredients:}
\begin{itemize}
\item 3 egg whites 
\item 1/2 cup of sugar 
\item 1 cup of really hard, frozen ice cream 
\item 1 big, thick, hard cookie 
\item Baking sheet 
\item Aluminum foil 
\item Hand mixer 
\item A grown up!!
\end{itemize} 

\subsection*{Directions:}
1. Have your grown up heat your oven to 500 degrees Fahrenheit. 
\newline
2. Cover the baking sheet with aluminum foil.
\newline
3. Put the egg whites into a bowl and use the mixer to beat them for about 
five minutes until they're stiff. 
\newline
4. Keep beating the egg whites while adding the sugar a little at a time until
they're fluffy and shiny. ("The name for this stuff is meringue!") 
\newline
5. Put your cookie on the baking sheet. 
\newline
6. Put your scoop of ice cream on top of the cookie. Make sure it doesn't hang
over the edge of the cookie. 
\newline
7. Completely cover the cookie and the ice cream with the meringue. 
\newline
8. Put it in the oven for three to five minutes until the meringue is 
a delicate, light brown. 
\newline
9. Take it out of the oven, put it on a plate, and eat up!

\end{document}